# The free path and the generation rate of a fast-moving electron interacting with a dielectric media

**Mykola Yelisieiev**[1 2]

Taras Shevchenko National University of Kyiv, Volodymyrska street 64, Kyiv, 01601, Ukraine

## Abstract

In the framework of macroscopic continuous medium approach, we studied the interaction between a fast-moving charged particle and dielectric or semiconducting media with low energy electrically active excitations. The excitations contribute to frequency dispersion of the media dielectric permittivity. Two types of processes induced by a moving charged particle are considered: electron-hole generation under interband transitions and excitation of polar optical phonons. For both processes we calculated and analyzed the time- and space-dependent electric potential generated by the charged particle, polarization of the media, energy losses of the particle and other important constituents of the interaction patterns. Obtained results can contribute to deeper understanding of the charged particle beams interaction with a semiconducting medium, as well as may be useful for versatile applications of charged beams.

[1] a.k.a. – Nicholas E. Eliseev

[2] Corresponding author, mykola.eliseev@gmail.com

## 1. Introduction

The problem of the interaction between fast charged particles and dielectric or semiconducting media is of particular interest for both fundamental standpoint and various applications. The fundamental aspects of the problem are related, for example, to the mechanisms of different type excitations in the medium, energy losses of the particles, etc. Applications of the problem include scanning electron microscopes using high-quality electron beams [1, 2, 3, 4], cathode luminescence effects [5], terahertz radiation from plasmon-polaritons excited by electrons [6, 7, 8], surface-plasmon resonance sensors [9], solid state charge particle detectors [10] and others [11, 12]. An important application of our results can be found in betavoltaics [13, 14]. In this technology the high-energy electrons pass through a dielectric media providing generation of electron-hole pairs, separation of which in *pn*-junctions gives rise to the voltaic effect. Energy cells based on this principle can be used in a number of applications [15-18].

In general, interaction of a charged particle with a media depends on atoms/ions composing the media and its particular structure. However, there exists examples, for which such a detailed knowledge is not necessary, instead a macroscopic continuous medium approach is applicable. Indeed, the moving charged particle induces time- and space-dependent electrostatic potential. At a distance $r$ from the particle trajectory the main contribution to this potential comes with frequencies of the order of $\omega \approx v_0/r$. Thus, electrically active excitations with low frequencies (and energies approximately equal to $\frac{\hbar v_0}{r}$), if they exist, can be produced at macroscopically large distances $r$ ($r >> a_0$ with $a_0$ being a lattice constant).

Such low energy excitations can be analyzed by using a macroscopic continuum media approach. This approach does not account for the impact of the lattice defects or separate atoms. It can be justified by the fact that we find and check a "characteristic length and frequency scales" in the process of our calculations. The scales are dependent on the speed of the charged particle.

Thus, implying a condition that the characteristic scale is considerably higher than the lattice constant, we obtain the lower limit of the particle speed:

$$v_0 \gg a\,\omega_{gap} = 1.67 \cdot 10^5 m/s. \qquad (1)$$

This means that the initial energy of the particle needs to be enough high. In this case the considered charge carrier is not influenced by the individual features of media, which can be treated macroscopically in frame of frequency-dependent dielectric permittivity. In this paper two models of electrical excitations will be studied.

For interband excitations induced by the fast electrons (e.g., cathode luminescence) we used the permittivity dispersion ε(ω) accounting transitions between electron energetic bands in a semiconductor media [12]:

$$\varepsilon' = \varepsilon_0 + \frac{A}{\chi^2}\left(2 - \sqrt{1+\chi} + \sqrt{1-\chi}\ \theta(1-\chi)\right), \qquad (2a)$$

$$\varepsilon'' = \frac{A}{\chi^2}\sqrt{\chi - 1}\ \theta(\chi - 1). \qquad (2b)$$

Here $\varepsilon'$ is the real part and $\varepsilon''$ is the imaginary one of the dielectric permittivity, $A$ is the oscillator force, and $\chi = \frac{\hbar\,\omega}{E_{gap}}$ is the dimensionless frequency, $E_{gap}$ is the energy band gap, $\theta$ is the Heaviside step function, $\varepsilon_0$ is a material constant.

For consideration of a polar material at the excitation of optical phonons we consider the permittivity in the form [15]:

$$\varepsilon_{phon} = \chi_\infty \frac{\omega_{lo}^2 - \omega^2 - 2i\gamma\omega}{\omega_{to}^2 - \omega^2 - 2i\gamma\omega}, \qquad (3)$$

where $\chi_\infty$ is the static permittivity of the media, and $\omega_{lo}, \omega_{to}$ are the frequencies of the longitudinal and transverse phonons, respectively [12]. The parameter $\gamma$ is the decay factor. The excitation of the polar medium can be characterized by the induced polarization, ***P***, which relates to the dielectric displacement ***D*** and the electric field ***E*** of the moving charge by the formula

$$\vec{D} = \vec{E} + 4\pi\,\vec{P}. \qquad (4)$$

## 2. Basic assumptions and equations

At first, we consider a particle with a charge $q$ moving with a speed $v$ in a medium with the dispersion law of the dielectric permittivity given by Eq. (2). The speed $v$ is assumed to be high, but sufficiently smaller than the light speed (nonrelativistic case), then we neglect magnetic fields around the moving charge particle and use electrostatic equations. In the first approximation the particle speed is considered to be constant (i.e., excited medium does not affect the particle motion), this allows to determine the electrical potential, the electric fields and the energy losses of the particle. Then, using the obtained energy losses we can write an approximate differential equation of particle deceleration. From this equation we determine the dependence of the particle speed on the coordinate along its motion. Because there are no magnetic fields, we use the Maxwell equation for the electric displacement in the differential form. Due to the frequency dispersion of media, we need to use the response convolution function [17]. By doing that it becomes possible to link the electric field and the displacement. So, the equations are:

$$\begin{cases} div(\vec{D}) = 4\,\pi\,q\;\delta(x - vt)\delta\;(y)\delta(z) \\ \vec{D}(x,y,z,t) = \int_{-\infty}^{0} \varepsilon(t-\tau)\vec{E}\;(x,y,z,\tau)d\tau \end{cases}. \tag{5}$$

In Eq. (5) the charge density of the moving particle of infinitesimal size is written by means of a Dirac delta-function. To make further calculations easier, we do a Fourier transform on time, and transfer to the Poisson equation for the scalar potential. It is possible to do because $rot\vec{E} = 0$. So, let us write the electric potential as follows:

$$\Delta\varphi(\omega,\vec{r}) = -\frac{4\,\pi\,q}{v\;\varepsilon(\omega)}\;e^{+i\omega\frac{x}{v}}\delta\;(y)\delta(z). \tag{6a}$$

The Fourier image of the Eq.(5) can be solved by using a Green function, and the general solution for the image of the potential is:

$$\varphi = \iiint_{-\infty}^{+\infty} G(x,\tilde{x},y,\tilde{y},z,\tilde{z})\ \frac{1}{v} e^{i\omega\frac{\tilde{x}}{v}}\delta\ (\tilde{y})\delta(\tilde{z})\ d\tilde{x}\ d\tilde{y}\ d\tilde{z}. \qquad (6b)$$

The boundary conditions for Eq.(5) are that the displacement or the potential decay to zero at infinity, and the Green function $G(x,\tilde{x},y,\tilde{y},z,\tilde{z})$ fulfills them.

### 3.1. The calculation of the electric potential

Implying the formula (6b) leads to the following form of the electric potential:

$$\varphi = B\left\{\int_0^1 \lambda(\chi)\ cos(\chi x')K_0\left(\chi\sqrt{y'^2+z'^2}\right)d\chi + \int_1^{+\infty} \mu(\chi)\ K_0\left(\chi\sqrt{y'^2+z'^2}\right)d\chi\right\}. \qquad (7a)$$

The following designations are used in the Eq.(7a):

$$\lambda(\chi) = \frac{1}{\varepsilon_0+\frac{A}{\chi^2}(2-\sqrt{1+\chi}+\sqrt{1-\chi}\ )}, \qquad (7b)$$

$$\mu(\chi) = \frac{\left(\varepsilon_0+\frac{A}{\chi^2}(2-\sqrt{1+\chi}\ )\right)cos(x'\chi)-\frac{A}{\chi^2}\sqrt{\chi-1}\ sin(x'\chi)}{\left(\varepsilon_0+\frac{A}{\chi^2}(2-\sqrt{1+\chi}\ )\right)^2+\frac{A^2}{\chi^4}(\chi-1)}. \qquad (7c)$$

And the dimensionless frequency

$$\chi = \frac{\hbar\,\omega}{E_{gap}}. \qquad (8a)$$

The dimensionless coordinates:

$$x' = \frac{x-vt}{l},\ y' = \frac{y}{l},\ z' = \frac{z}{l},\ \ r' = \sqrt{y'^2+z'^2}. \qquad (8b)$$

The characteristic length and frequency:

$$l = \frac{v}{\omega_{gap}},\ \omega_{gap} = \frac{E_{gap}}{\hbar} \qquad (8c)$$

$$B = \frac{q}{\pi\,v}\frac{E_{gap}}{\hbar} \qquad (8d)$$

The dependence of the induced potential (7) on the coordinates (8b) along the direction of the particle motion is shown in **Fig. 1**. As one can see from the figure, the potential oscillates for the negative values of the coordinate $x'$. The inset shows this behavior in more details. The contour map of the potential in coordinates $\{r', x'\}$ is presented in **Fig. 2**. Since the system is axially-symmetric, the radius-vector is the ordinate axis, perpendicular to the electron motion. The parameters used in these calculations are shown in **Tab. 1.** below.

**Table 1.** * The parameters of the media, used in our evaluations [16].

| Oscillator force $A$ | Material constant $\varepsilon_0$ | Band gap frequency $\omega_{gap}$ |
|---|---|---|
| 2.6 | 16 | $3 \times 10^{15}\ Hz$ |

*The material corresponding to these constants is InSb.

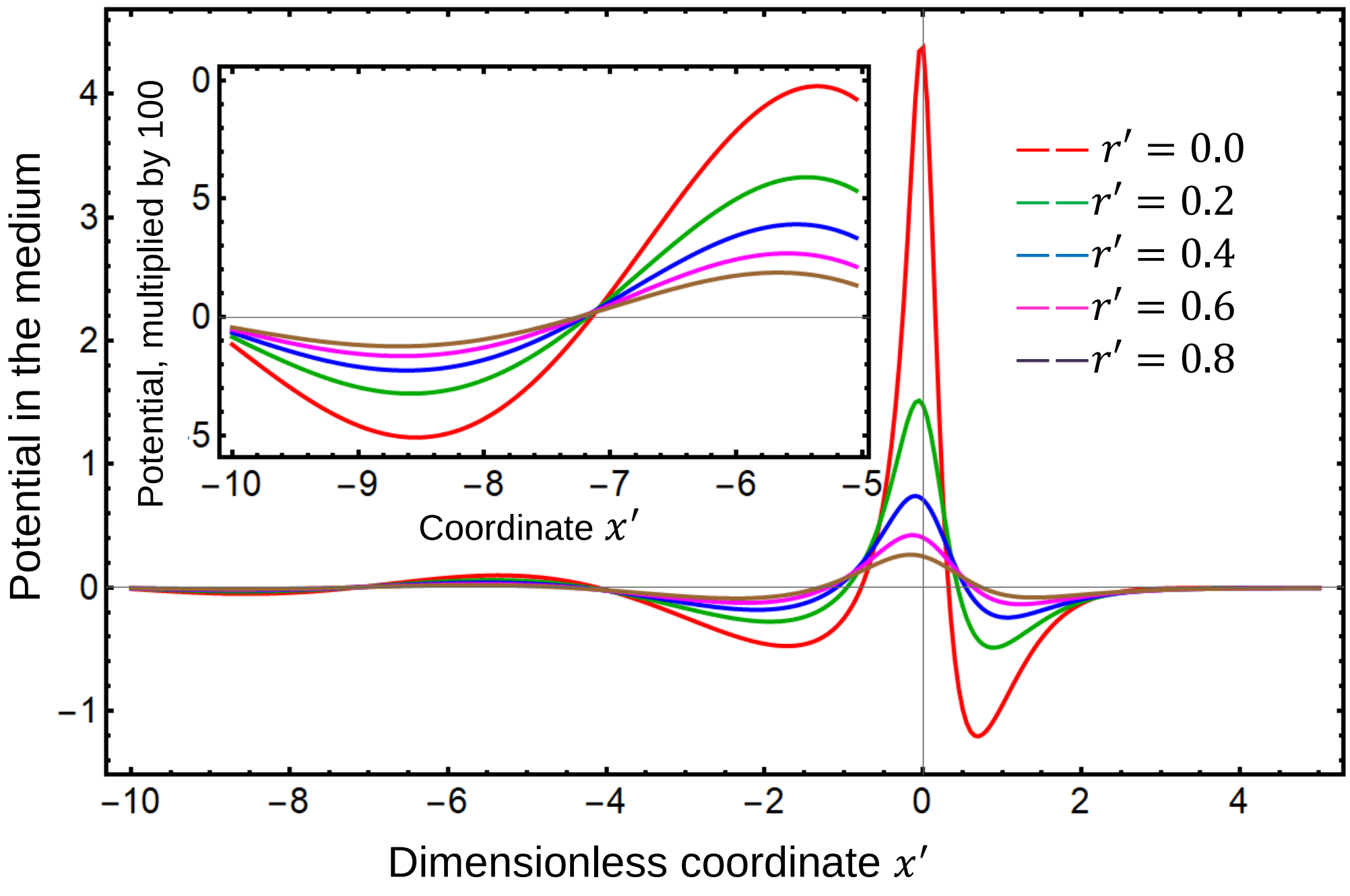

**Fig. 1**. The dependences of the induced potential versus longitudinal coordinate $x'$ for different radial distances $r'$.

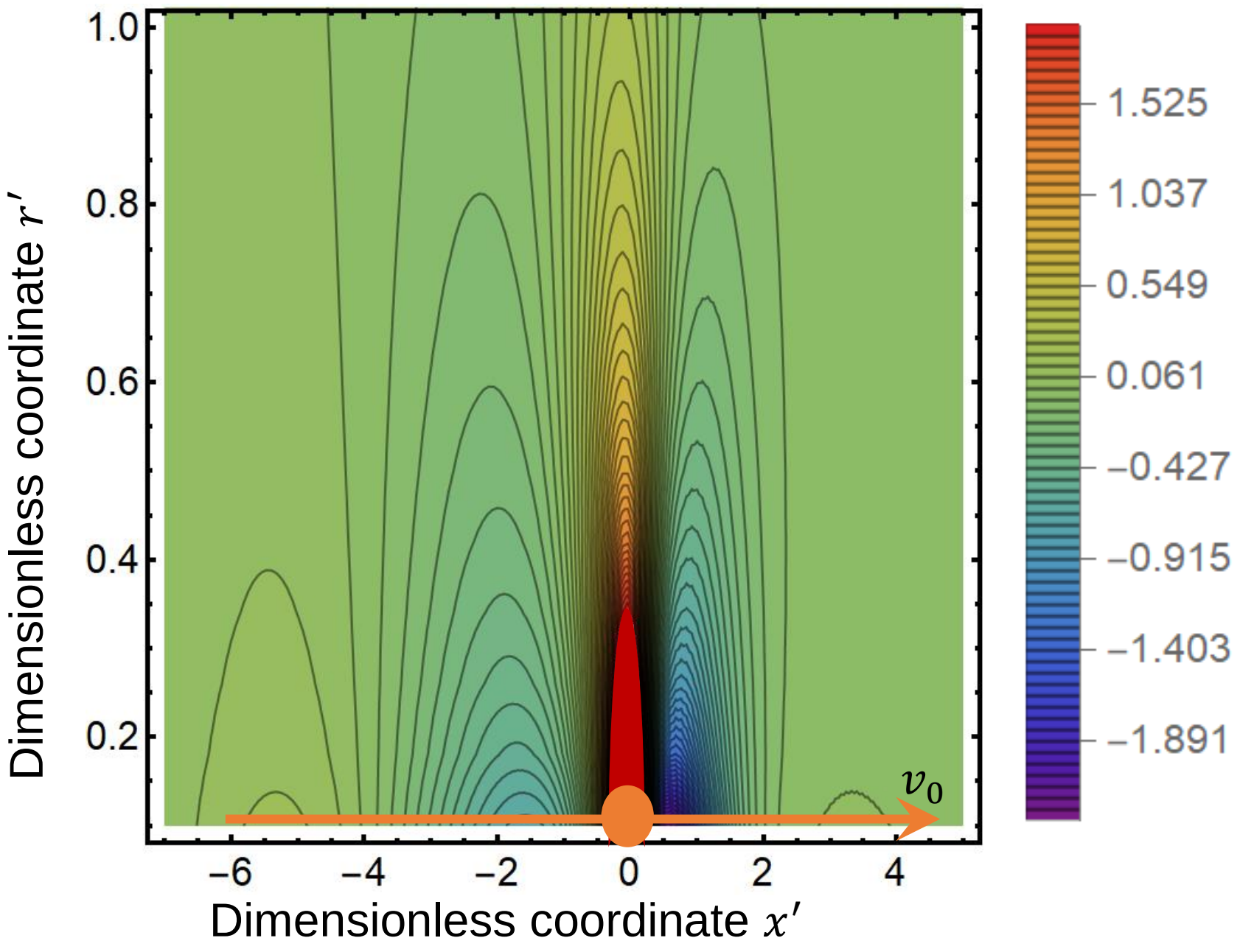

**Fig. 2**. The contour map of the electric potential given by Eq.(7). The system is axially-symmetric, so on the ordinate axis the value of the radius-vector, perpendicular to the motion of the electron, is shown. The orange arrow shows the direction of the electron movement.

The obtained electric potential is sign-alternating. It oscillates for the negative values of the longitudinal coordinate, and decays to zero in front of the particle. So, if one considers an electron beam in the same medium as a single particle, the property of the potential can lead to some peculiar effects in the beam. Namely, the density of the electrons in the beam can oscillate as well. There will be regions with increased and decreased density. The charged particles can gather in groups due to this effect. Our study cannot prove this mathematically, but there is a high probability that further studies will achieve this result.

Let us consider an electron beam, with electrons far enough from each other not to disturb their “neighbors” with the created electric filed. As one can see from **Fig. 2**, this distance must be not less than ten characteristic lengths (8c). If the electrons are distributed denser that the distance, their fields can affect their closest neighbors, and the effect of the density oscillation will occur.

As on can see from **Fig. 1**, the period of the electric potential oscillations is approximately six characteristic distances. We can say that if two electrons are situated closer than this distance, they influence each other. Thus, some correlations of the density in an electron beam can occur. This means that if an electron is located in a cube with an edge longer then the length of six characteristic distances, the density oscillations will not occur. Because in such case the electrons will not influence each other. So, we suggest a way of calculating the current density of a hypothetic beam, in which the electron correlations are still absent:

$$j = e\, v_0\, n \tag{9a}$$

The concentration $n$ can be estimated as one electron per approximately $10^3\, l^3$ volume. So, the current density can be written as follows

$$j = \frac{e\ \omega_{gap}^3}{10^3\, v_0^2}. \tag{9a}$$

### 3.2. The energy losses and the coordinate dependance of the particle speed

Let us find the energy losses in the media. To do that, we need to evaluate the field as the gradient of the potential (7) and the electric displacement as the response integral. Having the expressions for them, we will be able to find the energy. So, we can write the following formula, based on Ref.[17], page 306, formula (56.15):

$$\frac{d\mathcal{E}}{dt} = -\frac{1}{4\,\pi}\iiint_V\ \vec{E}\frac{\partial\vec{D}}{\partial t}dV. \tag{10}$$

These are the energy losses of the particle in a unit time (or the “power” of the losses). Hence, the particle energy losses per unit length can be written as:

$$d\mathcal{E}/dx = \int_{-\infty}^{\infty} dt \iint_{-\infty}^{+\infty} \rho_Q dydz, \tag{11}$$

where $\rho_Q = -\frac{1}{4\,\pi}\frac{\partial\vec{D}}{\partial t}\vec{E}$. The spatial density of the energy losses can be rewritten as follows:

$$\rho_{\mathcal{E}} = -\frac{1}{4\,\pi^2}\int_0^{+\infty} \omega \left|\vec{E}_\omega(\omega)\right|^2 \varepsilon'' d\omega = \int_0^{\infty} \rho_{\mathcal{E}}(\omega) d\omega. \qquad (12)$$

If we substitute the expression for $\vec{E}_\omega(\omega)$ into Eq. (12) we obtain the expression for the spatial density of the energy losses per unit frequency interval (*the spectral density*), $\rho_{\mathcal{E}}(\omega)$:

$$\rho_{\mathcal{E}}(\omega) = -\frac{1}{4\,\pi^2}\frac{q^2}{v_0^4}\frac{\omega^3\varepsilon''(\omega)}{|\varepsilon(\omega)|^2}\left[{K_0}^2\left(r_\perp\,\frac{\omega}{v}\right) + {K_1}^2\left(r_\perp\,\frac{\omega}{v}\right)\right]. \qquad (13)$$

In the expression (13) $r_\perp = \sqrt{y^2 + z^2}$ is the distance from the particle trajectory in the plane perpendicular to this trajectory. In the total energy losses, Eq. (11), the integrand, given by Eq. (13), diverges at $r_\perp \to 0$. To get a final value of the integral we perform the integration starting not from $r_\perp = 0$, but from a small distance *a*. Below we check that for the cut-off parameter *a* of the order of a few lattice constants, which is natural minimal distance of lattice atoms to the particle, the integral (11) is almost independent on *a*. This is shown in **Fig. 3**. We see that the integral term decreases with the increase of *a*, as a logarithmic function. For actual values of *a* above a few the lattice constant (5…10) Å, the integral changes very little. It proves the validity of the integral cut-off procedure.

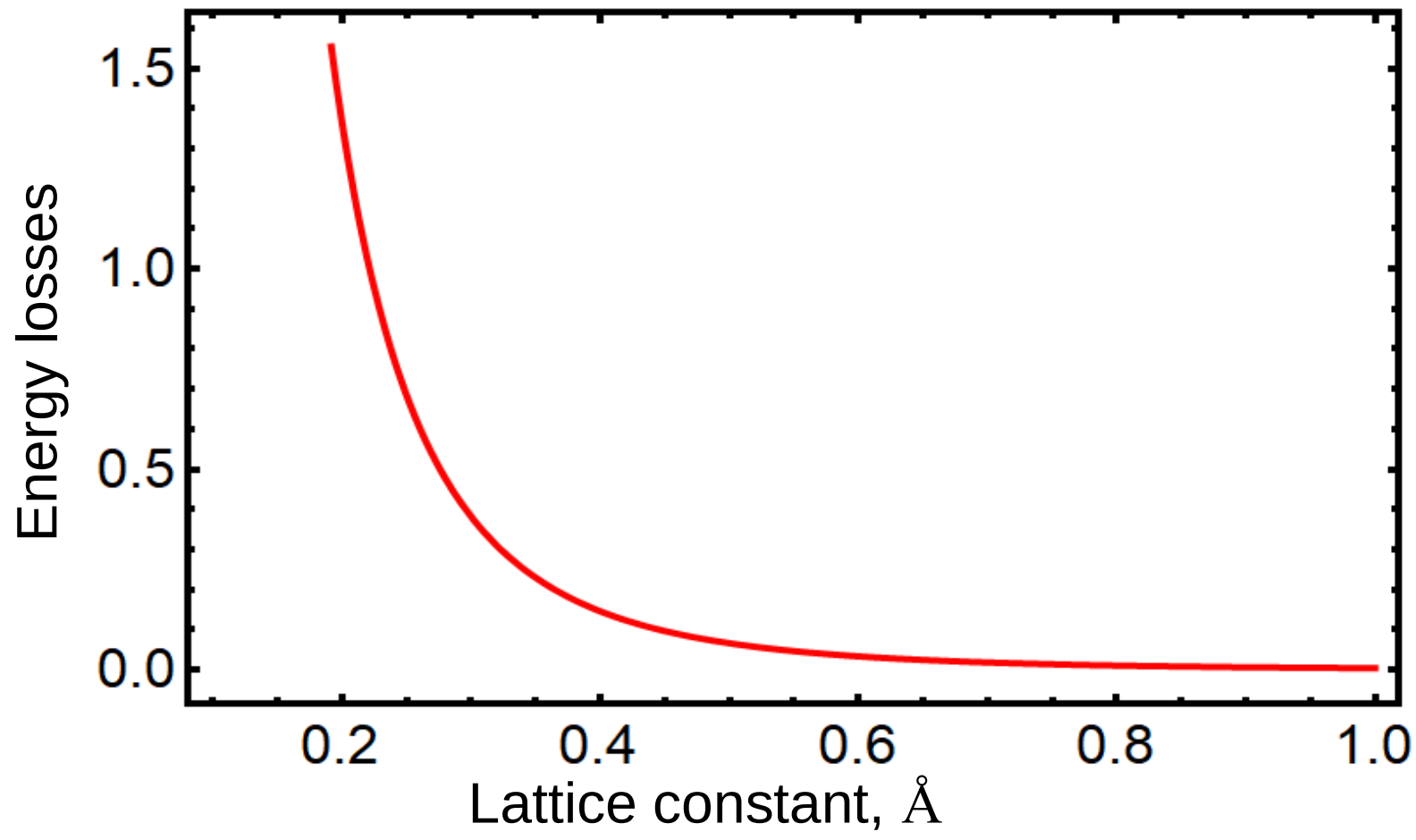

**Fig. 3.** The dependence of the energy losses per unit length on the cut-off parameter *a*.

We can equate the change of the kinetic energy $T = m\frac{v^2}{2}$ of the particle to the energy loss given by Eq. (11) and find the derivative of the particle velocity with respect to the coordinate $x$:

$$\frac{d}{dx}T = m\,v\,\frac{dv}{dx} = \frac{dE}{dx}. \tag{14}$$

Here $v$ is the speed of the particle. Now, the latter can be rewritten as

$$\frac{\Delta\mathcal{E}}{\Delta x} = -\frac{a^2}{4\pi\,A}\frac{q^2}{v_0^4\,\omega_{gap}{}^4}I. \tag{15}$$

Where we introduced the following parameter

$$I = \int_1^\infty \frac{\chi^5\sqrt{\chi-1}\left[K_1(a\,\chi)K_2(a\,\chi)-K_0{}^2(a\,\chi)\right]}{\kappa^2\chi^4+2\,\kappa\,\chi^2\left(2-\sqrt{1+\chi}\right)+(2\chi+4)-4\sqrt{1+\chi}}d\,\chi, \tag{16a}$$

$$\text{with } \kappa = \frac{\varepsilon_0}{A} = 6.153. \tag{16b}$$

The last value was obtained, using the parameters from **Tab. 1.** Then, we can solve the differential equation (14) and find the particle speed:

$$v = v_0\left[1-\frac{x}{x_{max}}\right]^{\frac{1}{6}}. \tag{17a}$$

Here $v_0$ is the initial speed of the particle. This result shows that the particle decelerates and stops after travelling the distance given by the expression

$$x_{max} = \frac{2\pi\,A\,m\,\omega_{gap}{}^4}{3\,a^2q^2I}v_0{}^6. \tag{17b}$$

All of these transformations and deliberations are possible only if characteristic frequency is significantly higher than the frequency of the interband transition:

$$\frac{v_0}{a} \gg \omega_{gap}, \quad v_0 \gg a\,\omega_{gap} = 1.67\cdot 10^5 m/s. \tag{17c}$$

So, we can write that $x_{max} > \frac{2\pi\, A\, m\, a^4\, \omega_{gap}{}^{10}}{3\, q^2 I}$. The schematic graphical form of the expression (17a) is shown in **Fig. 4.**

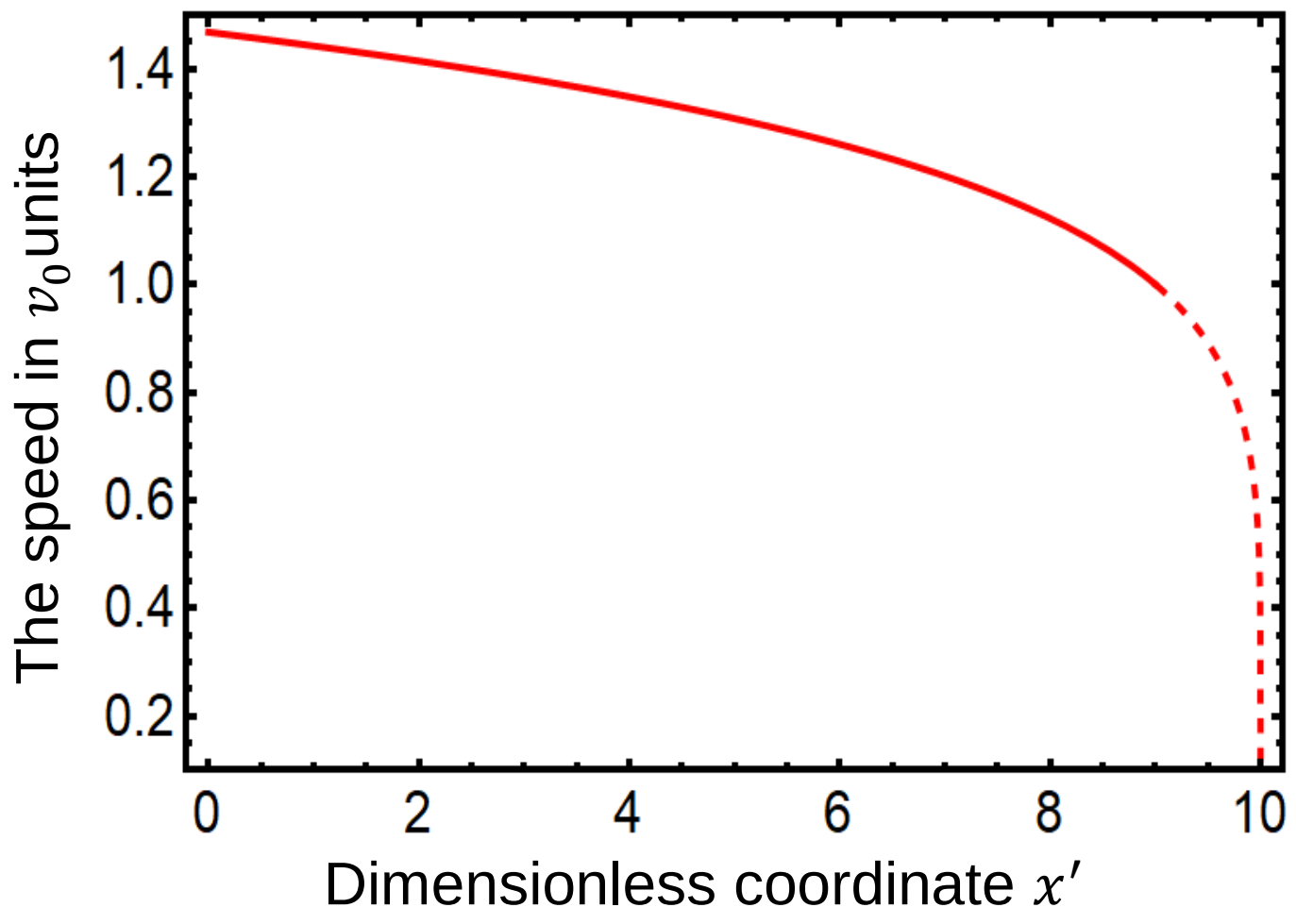

**Fig. 4.** The dependence of the particle speed on the coordinate $x'$ along its motion.

The result shown in **Fig. 4** shows that a charged particle in the vicinity of a dispersive media slows down and eventually stops. The formula (17a) shows an approximate expression for the maximal region of the particle motion. The speed is quasi-linear at first, then it rapidly decreases (see the dashed part on the graph). Our calculations can be considered trustworthy for the solid part of the curve, because of the condition (17c) on the value of the speed. The values of the speed represented by the dashed curve decrease too strongly from the initial value, thus they violate the condition (17c).

### 3.3. The carrier generation rate evaluation

Let us, knowing the losses density given by expression (13), evaluate the number of the charge carriers generated in a unit space per unit time. To be more precise, we need to know how many interband transitions one electron could cause with the energy it loses in the medium. We will consider such a transition with an energy equal to $\hbar\omega_t$. Let us write the following expressions:

$$\frac{d}{dt} n_{gen}(\omega_t)\, \hbar\omega_t = -\rho_{\mathcal{E}}(\omega_t)\,. \tag{18}$$

Here $n_{gen}$ is the concentration (a number per unit volume) of the interband transitions, $\omega_t$ is the transition frequency. The equation (18) follows from the energy conservation law. Since the frequency $\omega_t$ can have different values in the considered case, we need to integrate Eq. (18) over $\hbar\omega_t$. The result for the generation rate $G$ can be written as follows:

$$G(\tilde{r}_\perp) = \int_{\omega_{gap}}^{\infty} \frac{-\rho_{\mathcal{E}}(\omega_t)}{\hbar\,\omega_t} d\omega_t = \frac{1}{4\,\pi^2} \frac{q^2}{v_0^4} \int_{\omega_{gap}}^{\infty} \frac{\omega^2\, \varepsilon''(\omega)}{\hbar|\varepsilon(\omega)|^2} \left[ {K_0}^2 \left( r_\perp \frac{\omega}{v} \right) + {K_1}^2 \left( r_\perp \frac{\omega}{v} \right) \right] d\omega. \tag{19a}$$

After the conversion of the integral in (19a) to dimensionless parameters, we obtain the following expression:

$$G_{av}(\tilde{r}_\perp) = \int_1^{\infty} G_{av,\chi}\,(\chi, \tilde{r}_\perp) d\chi. \tag{19b}$$

Here the generation rate energy density can be written as follows:

$$G_{av,\chi}\,(\chi, \tilde{r}_\perp) = \frac{1}{4\,\pi^2\,\hbar\, v_0^4\, {\omega_{gap}}^3} \frac{q^2\; \chi^3\,\sqrt{\chi-1}\left[{K_0}^2(\tilde{r}_\perp\,\chi) + {K_1}^2(\tilde{r}_\perp\,\chi)\right]}{\kappa^2\chi^4 + 2\,\kappa\,\chi^2\left(2-\sqrt{1+\chi}\right) + (2\chi+4) - 4\sqrt{1+\chi}}. \tag{19c}$$

In Eq.(19c) $\chi = \frac{\mathcal{E}_{pair}}{\mathcal{E}_{gap}}$, and $\mathcal{E}_{pair}$ is the energy of the generated electron-hole pair. The dimensionless radius-vector absolute value is $\tilde{r}_\perp = \frac{r_\perp}{l}$. To illustrate the analytical expression (19b), a graphical image of the dependance of the generation rate dispersion $G_{av,\chi}\,(\chi, \tilde{r}_\perp)$ on the dimensionless frequency $\chi$ is shown below in **Fig. 5.**

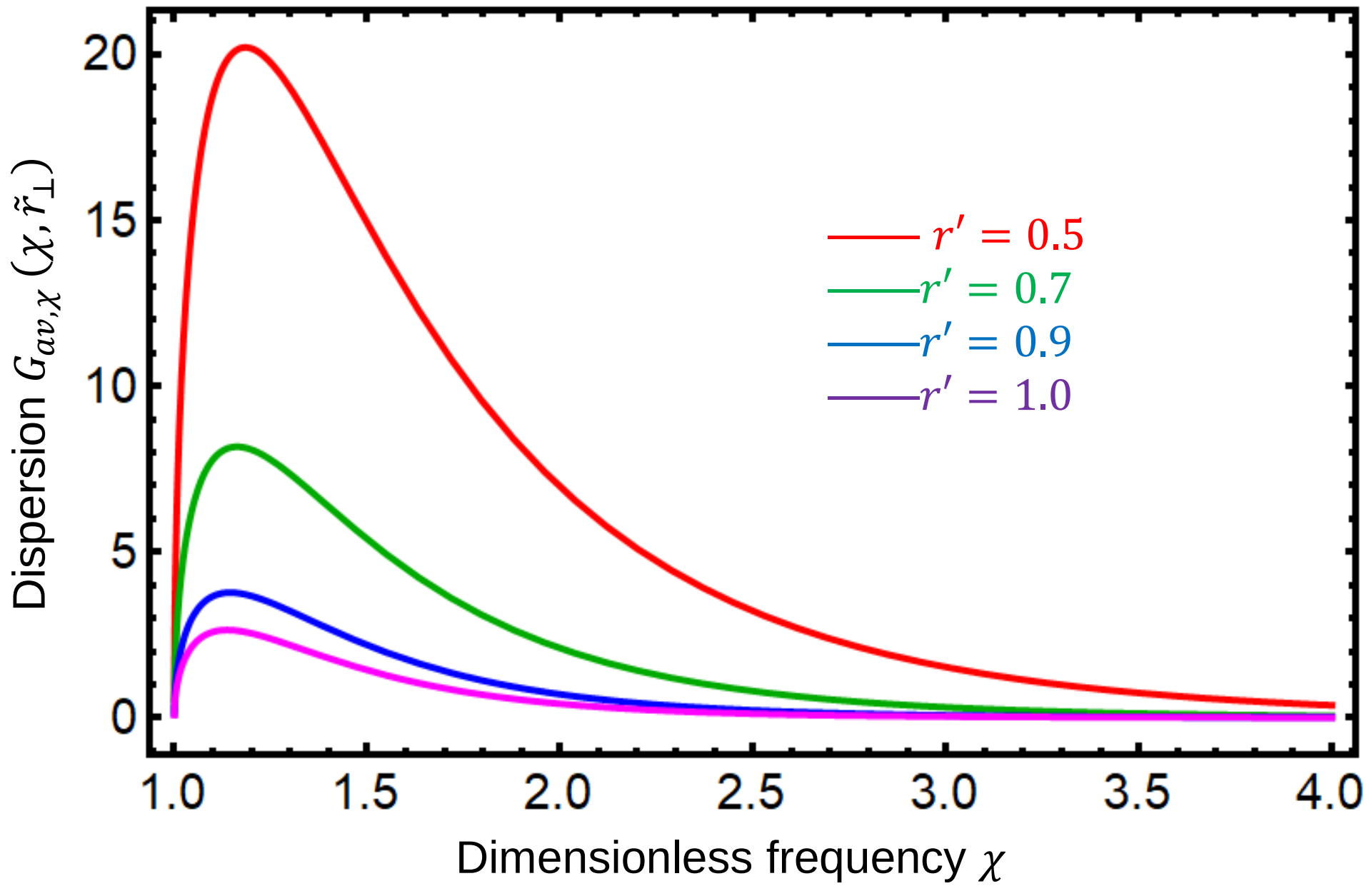

**Fig. 5.** The dependance of the generation rate dispersion $G_{av,\chi}(\chi, \tilde{r}_\perp)$ on the dimensionless frequency $\chi$ for different distances from the charge motion line listed in the legend.

As one can see from **Fig. 5**, the dispersion of the generation rate is significantly high for the transition energy close to the band-gap width. For much higher energies it is insignificant. This can be interpreted as the fact that the moving charge carrier generates only low-energy transitions. So, the macroscopic continuous medium approach is correct, because, as we can see from the figure, the mean energy of the generated transitions is very low. And, naturally, the speed increases with the decrease of the radius-vector absolute value.

Note, that the generation rate, and the distance of an electron penetration for the so-called betavoltaics (which is discussed in Ref. [18]) is critically important, because of the use of fast electrons with energies around 100 keV.

## Part II. A charged particle near a dielectric with optical phonons/vacuum boundary

### 4.1. Problem statement

Let us consider a charged particle, moving at a fixed distance from a boundary between two media. One of them is a dielectric without dispersion, the other is a dielectric with optic phonon dispersion. The speed of the particle is regarded constant. The schematic view of the considered system is shown in **Fig. 6.**

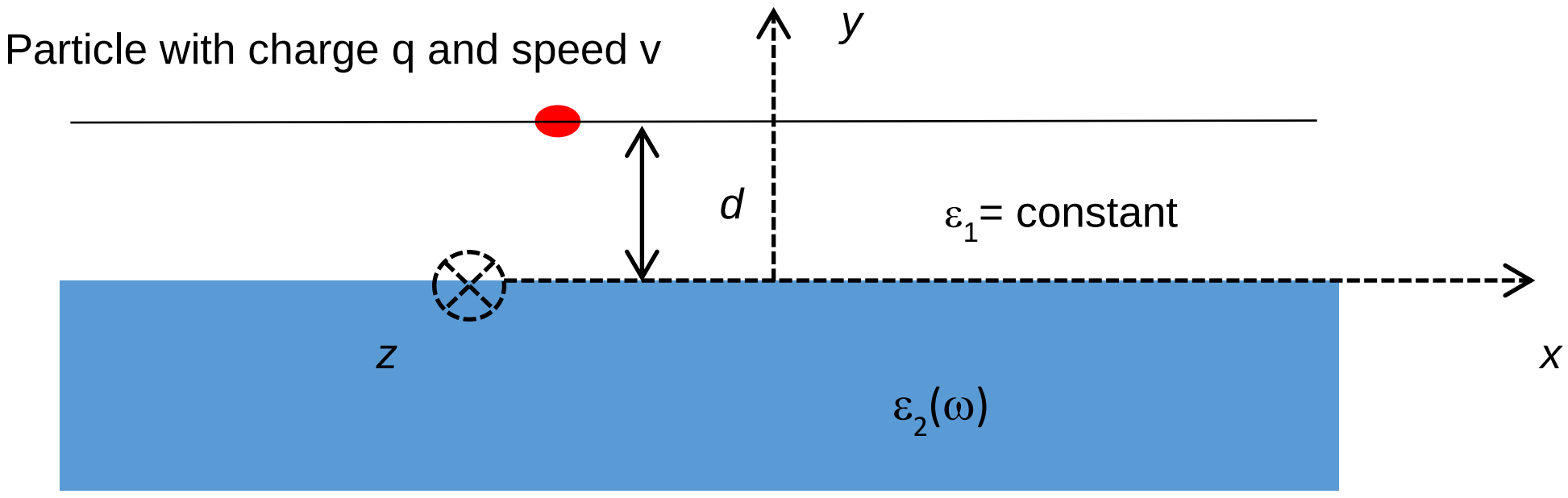

**Fig. 6.** A particle with a constant charge *q* and speed *v* moving along the boundary of two dielectrics at a distance *d* from it.

The permittivities of the media "1" and "2" are:

$$\varepsilon_1 = 1, \qquad \varepsilon_2 = \chi_\infty \frac{\omega_{lo}^2 - \omega^2 - 2i\gamma\omega}{\omega_{to}^2 - \omega^2 - 2i\gamma\omega}. \tag{20}$$

Let us find the potential, the electric field in both of the media, and the induction. The equations are considered the same as (5), with the following boundary conditions:

$$\begin{cases} \left(D_y^{(1)} - D_y^{(2)}\right)\Big|_{y=0} = 0 \\ \left(E_x^{(1)} - E_x^{(2)}\right)\Big|_{y=0} = 0 \end{cases}. \tag{21}$$

Next, we do a Fourier transform on time and the *x* coordinate, and a transition to the scalar electrostatic potential, because $rot(\vec{E}) = 0$. In result we obtain the following equations with boundary conditions:

$$\begin{cases} \Delta\varphi^{(1)}(\omega,\vec{r}) = -\frac{4\pi q}{v\,\varepsilon_1}\, e^{+i\omega\frac{x}{v}}\delta\,(y-d)\delta(z), \qquad y>0 \\ \varepsilon_2(\omega)\Delta\varphi^{(2)}(\omega,\vec{r}) = 0, \qquad y<0 \\ \varphi^{(1)}\Big|_{y=0} = \varphi^{(2)}\Big|_{y=0} \\ \varepsilon_1 \frac{\partial\varphi^{(1)}}{\partial y}\Big|_{y=0} = \varepsilon_2 \frac{\partial\varphi^{(2)}}{\partial y}\Big|_{y=0} \end{cases} . \qquad (22)$$

### 4.2. The electric potential and the field

The solution of Eqs.(22) was found by the means of a Green function, and has the form:

$$\varphi_1 = \frac{2q}{\pi d}\Big\{\int_0^{+\infty} cos(\tilde{\omega}x')\left[K_0\big(\tilde{\omega}\tilde{B}_-\big) + K_0\big(\tilde{\omega}\tilde{B}_+\big)\right]d\tilde{\omega} + \int_0^{\infty} d\tilde{\omega}\big[-2(\cos(\tilde{\omega}x')\,\tau(\tilde{\omega}) + sin(\tilde{\omega}x')\sigma(\tilde{\omega}))K_0\big(\tilde{\omega}\tilde{B}_+\big)\big]d\tilde{\omega}\Big\}. \qquad (23)$$

The Eq. (23) is written for $y > 0$, in the upper half-space. In the lower half-space (for $y < 0$) the potential has the form:

$$\varphi_2 = \frac{4q}{\pi d}\int_0^{+\infty}\big(cos(\tilde{\omega}x') - \{cos(\tilde{\omega}x')\tau(\tilde{\omega}) + sin(\tilde{\omega}x')\sigma(\tilde{\omega})\}\big)\; K_0(\,\tilde{\omega}\tilde{B}_-)d\tilde{\omega}. \qquad (24)$$

In Eq. (23-24) the following dimensionless factors were used:

$$\tau(\tilde{\omega}) = \frac{\chi_\infty\Big(\big(\tilde{\omega}_{lo}^2-\tilde{\omega}^2\big)^2+4(\tilde{\gamma}\tilde{\omega})^2\Big)+\Big(\big(\tilde{\omega}_{lo}^2-\tilde{\omega}^2\big)\big(\tilde{\omega}_{to}^2-\tilde{\omega}^2\big)+4(\tilde{\gamma}\tilde{\omega})^2\Big)}{\Big[\big(\tilde{\omega}_{to}^2-\tilde{\omega}^2\big)+\chi_\infty\big(\tilde{\omega}_{lo}^2-\tilde{\omega}^2\big)\Big]^2+4\,\tilde{\gamma}^2\tilde{\omega}^2(1+\chi_\infty)^2}, \qquad (25a)$$

$$\sigma(\tilde{\omega}) = \frac{\chi_\infty 2\tilde{\gamma}\tilde{\omega}\big(\tilde{\omega}_{to}^2-\tilde{\omega}_{lo}{}^2\big)}{\Big[\big(\tilde{\omega}_{to}^2-\tilde{\omega}^2\big)+\chi_\infty\big(\tilde{\omega}_{lo}^2-\tilde{\omega}^2\big)\Big]^2+4\,\tilde{\gamma}^2\tilde{\omega}^2(1+\chi_\infty)^2}. \qquad (25b)$$

In Eqs.(23)-(25) the following dimensionless parameters were introduced:

$$\tilde{\omega}_{to} = \frac{\omega_{to}}{\omega_v},\ \tilde{\omega}_{lo} = \frac{\omega_{lo}}{\omega_v},\ \tilde{\gamma} = \frac{\gamma}{\omega_v}, \qquad (26a)$$

$$\tilde{B}_\pm = \sqrt{(y' \pm 1)^2 + (z')^2}, \qquad (26b)$$

$$x' = \frac{x-vt}{d},\ \ y' = \frac{y}{d},\ \ z' = \frac{z}{d}, \qquad (26c)$$

$$\widetilde{\omega} = \frac{\omega}{v} d, \quad \omega_v = \frac{v}{d}. \tag{26d}$$

The coordinate dependence of the potential represented by Eqs. (23) and (24) is shown in **Fig. 7**. The curves were plotted for the parameters listed in **Tab. 2**.

**Table 2.** * The dimensionless parameters values, used in our calculations.

| $\tilde{\gamma}$ | $\widetilde{\omega}_{lo}$ | $\widetilde{\omega}_{to}$ | $d$ | $\chi_\infty$ |
|---|---|---|---|---|
| 0.01 | 1.2 | 1 | $10^{-4}$ | 2.56 |

*The material corresponding to these constants is CaAs.

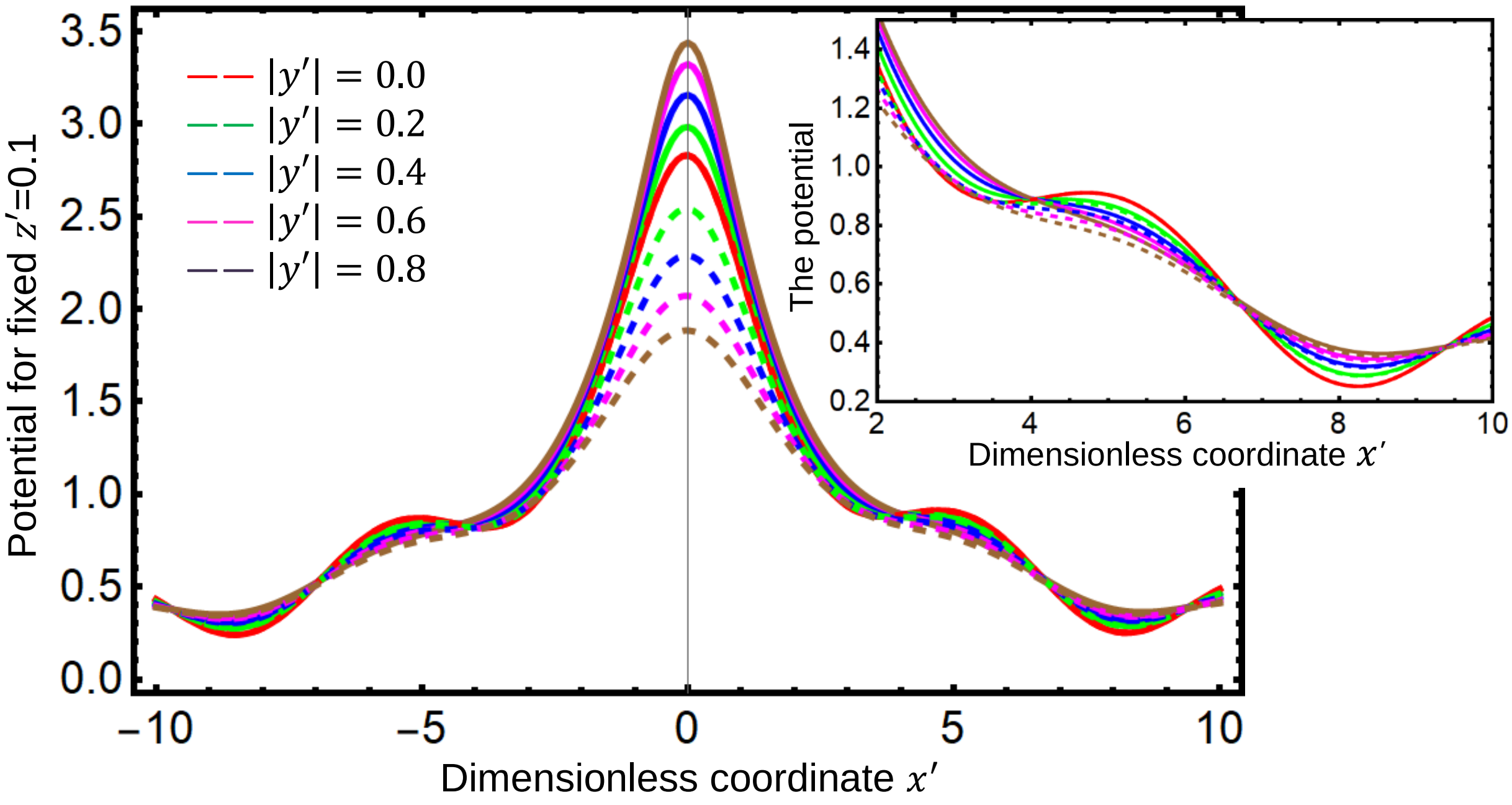

**Fig. 7.** The dependence of the electrostatic potential on the dimensionless coordinate $x'$ for fixed $z'$ and different $y'$ (different solid and dashed curves). The solid curves correspond to the upper half-space, the dashed curves – to the lower half-space of the system. The inset shows the oscillations of the potential in more details.

The electric potential oscillates for the positive values of the dimensionless coordinate, as is shown in the inset of **Fig. 7**. For the negative values it oscillates as well. So, the changes of density in an electron beam will occur in this case as well. These oscillations occur for molecules as well, as is shown in [19].

From the expressions (23) and (24) one can obtain the formulas for the electric field and the polarization in the medium. They are shown below:

$$E_x = \frac{4}{\pi}\frac{q}{d^2}\frac{\chi_\infty}{\varepsilon_1}\int_0^{+\infty}\tilde{\omega}\left[-\frac{2}{\varepsilon_1}sin(\tilde{\omega}x') - sin(\tilde{\omega}x')\tau(\tilde{\omega}) + cos(\tilde{\omega}x')\sigma(\tilde{\omega})\right]\ K_0\left(\tilde{\omega}\tilde{B}_-\right)d\tilde{\omega}. \qquad (27a)$$

The $x$-component of the polarization can be written as follows:

$$P_x = \frac{q}{\pi^2 d^2}\frac{\chi_\infty}{\varepsilon_1}\int_0^{+\infty}\tilde{\omega}\left\{[\varepsilon'(\tilde{\omega}) - 1]\left\{-\frac{2}{\varepsilon_1}sin(\tilde{\omega}x') - sin(\tilde{\omega}x')\tau(\tilde{\omega}) + cos(\tilde{\omega}x')\sigma(\tilde{\omega})\right\} - \varepsilon''(\tilde{\omega})\left\{\frac{2}{\varepsilon_1}cos(\tilde{\omega}x') + sin(\tilde{\omega}x')\sigma(\tilde{\omega}) + cos(\tilde{\omega}x')\tau(\tilde{\omega})\right\}\right\}\ K_0\left(\tilde{\omega}\tilde{B}_-\right)d\tilde{\omega}. \qquad (28)$$

Here the imaginary $\varepsilon''(\tilde{\omega})$ and the real $\varepsilon'(\tilde{\omega})$ parts of the permittivity are derived from Eq. (3). Let us analyze a contour plot of the longitudinal (to the motion of the particle) component of the polarization given by Eq. (28), and the field, as from Eq. (27). They are shown on **Fig. 8 (a**) and **(b)**, respectively.

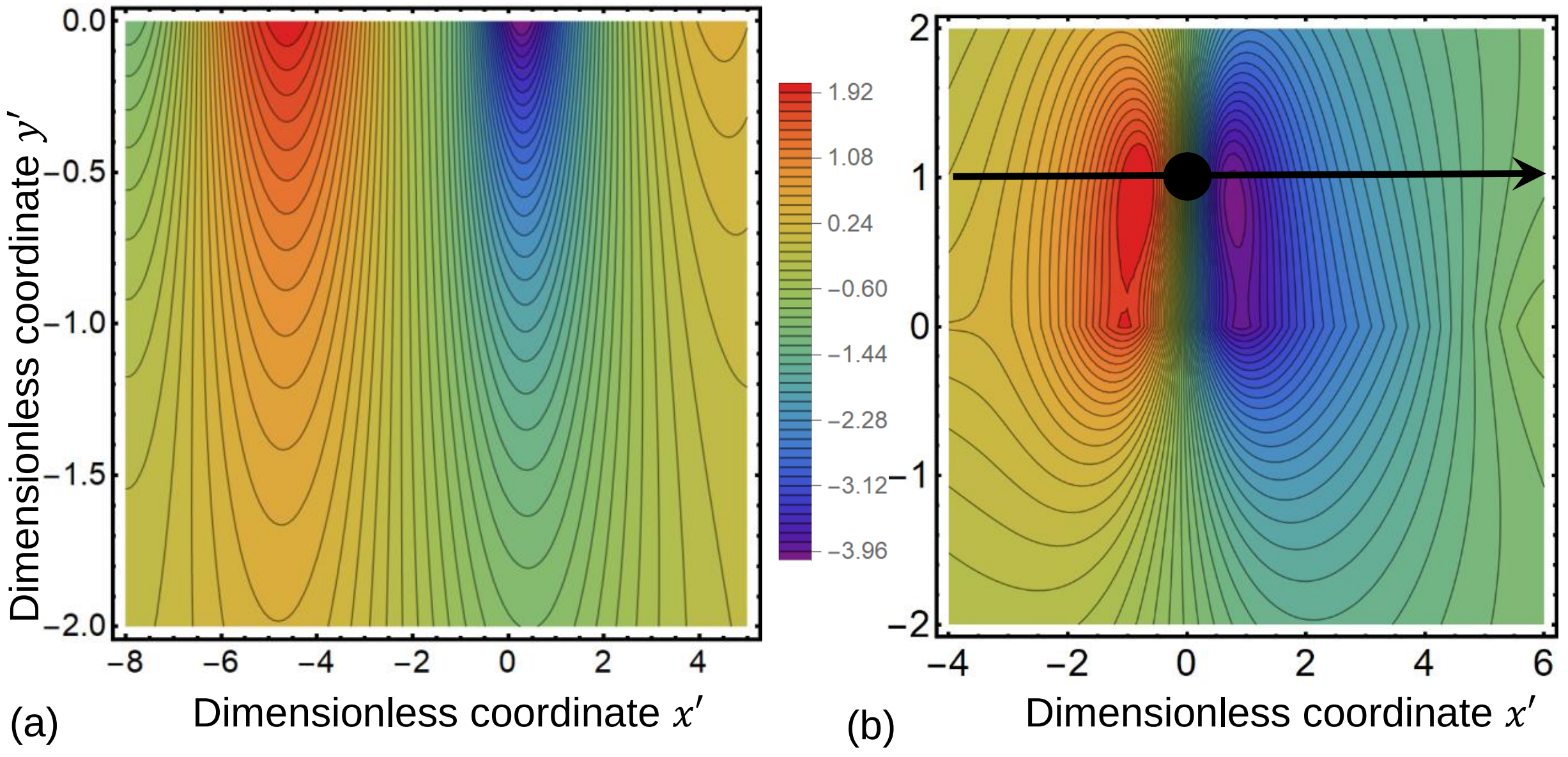

**Fig. 8.** (a) The space distribution of the x-component of the polarization in the dispersive media (28a). (b) The space distribution of the electric field. The black arrow shows the trajectory of the charged particle, the point – its position.

As one can see, the electron induces two regions with a significant value of the polarization – one five characteristic distances behind it, and one right in front of it. The field has a similar structure – two regions with significant in value and opposite in sign x-components, one two characteristic distances in front, other two of them behind. If we consider an electron beam in these conditions, we will find that the minimal critical concentration corresponding to the density correlations is equal to $\frac{1}{5^3 d^3}$.

## Conclusions

In this work we considered the interaction a fast-moving charged particle with low-energy electrically active excitations of a dielectric/semiconducting medium. The long-range character of the potential induced by the charged particle facilitates analysis of the low-energy excitations by applying the macroscopic description of dielectric/semiconducting medium. Such a macroscopic description can be based on characterization of the medium by a dielectric permittivity. Frequency dependence of the dielectric permittivity represents low-energy and electrically active excitations of analyzed medium.

Two particular types of the low-energy excitations were studied: (i) interband excitations in a semiconductor, (ii) optical phonon excitations in a polar dielectric. Then, we considered a charged particle moving through an infinite medium, and one moving near the boundary of two dielectric media. We found that for every excitation type and geometry the electric potential induced by the particle, as well as the medium polarization, oscillates with the coordinate along the particle trajectory. Excitations of the medium give rise to energy losses of the moving particle and its deceleration. We found the approximate coordinate dependence of the particle speed and its path length. We determined the spatial patterns of the induced potential, generation rate of the electron-hole pairs under interband excitation and polarization of polar medium.

For the excitation type (i) we found an analytical formula for the spectral density of the generation rate of electron-hole pairs induced by the fast-moving charged particle. We proved that the spectral density has significant amplitude only for the frequencies close to the bandgap and rapidly increases in the vicinity of the charge trajectory.

For the excitation type (ii) we found that the moving charged particle creates complex patterns of the electric field and polarization: two regions attendant the particle have significant amplitudes of the field and the polarization, signs of the field and the polarization are opposite in these regions. The moving patterns of the polarization induce complex lattice vibrations in the particle trace.

We suggest that the obtained results can contribute to deeper understanding of physics of interaction of charged particles with dielectric/semiconducting media, as well as may be useful for numerous devices and technologies using charged beam-medium interaction.

**Acknowledgements.** I express the deepest gratitude to my scientific supervisor Prof. Dr. V.A. Kochelap for the problem statement, guidance, help with result analysis and permanent stimulating discussions.

[19] M. S. Tomassone and A. Widom, Electronic friction forces on molecules moving near metals, *Phys. Rev. B* **56**, pp 4938–4943, (1997).

## Supplement 1

Let us consider a moving with a constant speed charge carrier. Let it move in an unlimited medium with dispersion. The equations are as follows:

$$\begin{cases} div(\vec{D}) = 4\,\pi\,q\,\delta(x - vt)\delta\,(y)\delta(z) \\ \vec{D}(x, y, z, t) = \int_{-\infty}^{0} \varepsilon(t - \tau)\vec{E}\,(x, y, z, \tau)d\tau \end{cases} \quad (1)$$

Let us do a Fourier transform on the variable *t*:

$$\vec{D}(x, y, z, \omega) = \varepsilon(\omega)\,\vec{E}(x, y, z, \omega) \quad (2a)$$

$$div\vec{D}(\,\omega) = \int_{-\infty}^{+\infty} 4\,\pi\,q\,\delta(x - vt)\delta\,(y)\delta(z)e^{i\omega t}dt = \frac{4\,\pi\,q}{v}\,e^{+i\omega\frac{x}{v}}\delta\,(y)\delta(z) \quad (2b)$$

After implying the scalar potential definition one can obtain, that

$$\Delta\varphi(\omega, \vec{r}) = -\frac{4\,\pi\,q}{v\,\varepsilon(\omega)}\,e^{+i\omega\frac{x}{v}}\delta\,(y)\delta(z) \quad (3)$$

The permittivity can be written in the following form:

$$\varepsilon(\omega) = \varepsilon' + i\,\varepsilon'' \quad (4a)$$

$$\begin{matrix} \varepsilon' = \varepsilon_0 + \frac{A}{\chi^2}\left(2 - \sqrt{1+\chi} + \sqrt{1-\chi}\;\theta(1-\chi)\right) \\ \varepsilon'' = \frac{A}{\chi^2}\sqrt{\chi - 1}\;\theta(\chi - 1) \end{matrix} \quad (4b)$$

In (4b) $A = 2.6, \chi = \frac{\hbar\,\omega}{E_{gap}}$, $\theta$ is the Heaviside theta function.

$$G(\omega, x, \tilde{x}, y, \tilde{y}, z, \tilde{z}) = \frac{q}{v\,\varepsilon(\omega)}\frac{1}{\sqrt{(x-\tilde{x})^2 + (y-\tilde{y})^2 + (z-\tilde{z})^2}} \quad (5)$$

And the solution can be evaluated in the following way:

$$\varphi = \iiint_{-\infty}^{+\infty} G(\omega, x, \tilde{x}, y, \tilde{y}, z, \tilde{z})\,\frac{1}{v}e^{i\omega\frac{\tilde{x}}{v}}\delta\,(\tilde{y})\delta(\tilde{z})\;d\tilde{x}\;d\tilde{y}\;d\tilde{z} =$$

$$\int_{-\infty}^{+\infty} G(\omega, x, \tilde{x}, y, 0, z, 0)\,\frac{1}{v}e^{i\omega\frac{\tilde{x}}{v}}d\tilde{x} = \int_{-\infty}^{+\infty} \frac{q}{v\,\varepsilon(\omega)}\frac{1}{\sqrt{(x-\tilde{x})^2 + y^2 + z^2}}\,\frac{1}{v}e^{i\omega\frac{\tilde{x}}{v}}d\tilde{x} \quad (6)$$

Now let us transform $\frac{1}{\varepsilon(\omega)}$ to separate the real and the imaginary terms:

$$\alpha(\omega) = \frac{1}{\varepsilon(\omega)} = \frac{\varepsilon' - i\,\varepsilon''}{\varepsilon'^2 + \varepsilon''^2} \quad (6a)$$

But the approximate formulae (4) was written for the positive values of $\chi$. So, we need to modify (4) for the negative range of the frequencies. As is written by Landau, the permittivity has the following symmetry properties:

$$\begin{matrix} \varepsilon'(-\omega) = \varepsilon'(\omega) \\ \varepsilon''(-\omega) = -\varepsilon''(\omega) \end{matrix} \quad (6b)$$

So, we need to modify (4a-b) accordingly to (6b):

$$\varepsilon' = \varepsilon_0 + \frac{A}{\chi^2}\left(2 - \sqrt{1+\chi}\,\theta(\chi) - \sqrt{1-\chi}\,\theta(-\chi) + \sqrt{1-|\chi|}\;\theta(1-\chi^2)\right)$$
$$\varepsilon'' = \frac{A}{\chi^2}\sqrt{\chi-1}\;\theta(\chi)\,\theta(\chi-1) - \frac{A}{\chi^2}\sqrt{-\chi-1}\,\theta(-\chi)\,\theta(-\chi-1) \tag{6c}$$

As one can see, the real part of (6a) is even, and the imaginary – uneven. So, the imaginary part of the integral (6) is identically equal to zero due to the integrand being uneven.

The integral (6) can be transformed and evaluated in the following way:

$\int_0^\infty \frac{cos(k\,ב)}{\sqrt{k^2+\vartheta^2}}\; e^{-א\sqrt{k^2+\vartheta^2}}dk = K_0\left(\vartheta\,\sqrt{ב^2 + א^2}\right)$ For such transformation, let us do the following variables interchange: $x - \tilde{x} = \mathfrak{E}$. So, our integral can be written in the following form:

$$\int_{-\infty}^{+\infty} \frac{q}{A(\omega)\sqrt{\mathfrak{E}^2+\vartheta^2}} e^{i\omega\frac{x}{v}} e^{i\omega\frac{\mathfrak{E}}{v}} d\mathfrak{E} = \int_{-\infty}^{+\infty} \frac{q\left[\cos\left(\omega\frac{\mathfrak{E}}{v}\right) + i\sin\left(\omega\frac{\mathfrak{E}}{v}\right)\right]}{A(\omega)\sqrt{\mathfrak{E}^2+\vartheta^2}} e^{i\omega\frac{x}{v}} d\mathfrak{E} = 2\int_0^{+\infty} \frac{q\;\cos\left(\omega\frac{\mathfrak{E}}{v}\right)}{A(\omega)\sqrt{\mathfrak{E}^2+\vartheta^2}} e^{i\omega\frac{x}{v}} d\mathfrak{E} = \frac{2\,q}{A(\omega)} e^{i\omega\frac{x}{v}} K_0\left(\vartheta\;\frac{|\omega|}{v}\right) \tag{7}$$

$$\text{So, } \varphi = \frac{q}{v\;\varepsilon(\omega)} e^{i\omega\frac{x}{v}} K_0\left(\sqrt{y^2+z^2}\;\frac{|\omega|}{v}\right) \tag{8}$$

Now, let us do a reverse Fourier transform on $\omega$:

$$\varphi(t,\vec{r}) = \frac{1}{2\,\pi}\int_{-\infty}^{+\infty} e^{-i\omega t}\varphi\,(\omega,\vec{r})d\omega = \frac{q}{2\,\pi\,v}\int_{-\infty}^{+\infty} \alpha(\omega) e^{-i\omega t} e^{i\omega\frac{x}{v}} K_0\left(\sqrt{y^2+z^2}\;\frac{|\omega|}{v}\right) d\omega \tag{9}$$

Let us introduce the following dimensionless variables:

$$\widetilde{\omega} = \chi = \frac{\hbar\,\omega}{E_{gap}} \tag{10}$$

$$x' = \frac{x - vt}{l}$$
$$y' = \frac{y}{l} \tag{11}$$
$$z' = \frac{z}{l}$$

$$l = \frac{v}{\omega_{gap}}$$
$$\omega_{gap} = \frac{E_{gap}}{\hbar} \tag{12}$$

After a transfer to the said dimensionless parameters and some simple algebraical transformations one can obtain, that

$$\alpha(\omega) = \frac{\varepsilon' - i\,\varepsilon''}{\varepsilon'^2 + \varepsilon''^2} \tag{13a}$$

$$Re\left[\alpha(\omega)e^{-i\omega t}\right] = \frac{\varepsilon'}{\varepsilon'^2+\varepsilon''^2} cos(\omega t) - \frac{\varepsilon''}{\varepsilon'^2+\varepsilon''^2} sin(\omega t) \tag{13b}$$

$$\varphi = \frac{q}{2\,\pi\,v}\frac{E_{gap}}{\hbar}\left[\int_{-\infty}^{+\infty} \frac{\varepsilon'}{\varepsilon'^2+\varepsilon''^2} K_0\left(\sqrt{y'^2+z'^2}\;|\chi|\right) cos(x'\chi)\,d\chi - \int_{-\infty}^{+\infty} \frac{\varepsilon''}{\varepsilon'^2+\varepsilon''^2} K_0\left(\sqrt{y'^2+z'^2}\;|\chi|\right) sin(x'\chi)\,d\chi\right] \tag{13c}$$

$$\frac{\varepsilon'}{\varepsilon'^2+\varepsilon''^2} = \frac{\varepsilon_0 + \frac{A}{\chi^2}\left(2 - \sqrt{1+\chi}\,\theta(\chi) - \sqrt{1-\chi}\,\theta(-\chi) + \sqrt{1-|\chi|}\;\theta(1-\chi^2)\right)}{\beta(\chi)} \tag{13d}$$

$$\frac{\varepsilon''}{\varepsilon'^2+\varepsilon''^2} = \frac{\frac{A}{\chi^2}\sqrt{\chi-1}\ \ \theta(\chi)\,\theta(\chi-1) - \frac{A}{\chi^2}\sqrt{-\chi-1}\,\theta(-\chi)\,\theta(-\chi-1)}{\beta(\chi)} \quad (13e)$$

Where

$$\beta(\chi) = \left(\varepsilon_0 + \frac{A}{\chi^2}\left(2 - \sqrt{1+\chi}\,\theta(\chi) - \sqrt{1-\chi}\,\theta(-\chi) + \sqrt{1-|\chi|}\ \ \theta(1-\chi^2)\right)\right)^2 + \frac{A^2}{\chi^4}(\chi-1)\ \ \theta(\chi)\,\theta(\chi-1) + \frac{A^2}{\chi^4}(-\chi-1)\,\theta(-\chi)\,\theta(-\chi-1) \quad (13f)$$

So, the integral (13c) can be re-written in the following way:

$$\varphi = B\int_{-\infty}^{+\infty}\left[\frac{\varepsilon' cos(x'\chi)}{\varepsilon'^2+\varepsilon''^2} - \frac{\varepsilon'' sin(x'\chi)}{\varepsilon'^2+\varepsilon''^2}\right]K_0\left(\sqrt{y'^2+z'^2}\ |\chi|\right)d\chi =$$

$$2B\int_0^{+\infty}\left[\frac{\varepsilon_0+\frac{A}{\chi^2}\left(2-\sqrt{1+\chi}+\sqrt{1-|\chi|}\ \ \theta(1-\chi^2)\right)}{\left(\varepsilon_0+\frac{A}{\chi^2}\left(2-\sqrt{1+\chi}+\sqrt{1-|\chi|}\ \ \theta(1-\chi^2)\right)\right)^2+\frac{A^2}{\chi^4}(\chi-1)\ \ \theta(\chi-1)}cos(x'\chi) - \frac{\frac{A}{\chi^2}\sqrt{\chi-1}\ \ \theta(\chi-1)}{\left(\varepsilon_0+\frac{A}{\chi^2}\left(2-\sqrt{1+\chi}+\sqrt{1-|\chi|}\ \ \theta(1-\chi^2)\right)\right)^2+\frac{A^2}{\chi^4}(\chi-1)\ \ \theta(\chi-1)}sin(x'\chi)\right]K_0\left(\sqrt{y'^2+z'^2}\ \chi\right)d\chi \quad (14a)$$

So, finally the electrostatic potential can be written as follows:

$$\varphi = B\left\{\int_0^1 \frac{1}{\varepsilon_0+\frac{A}{\chi^2}\left(2-\sqrt{1+\chi}+\sqrt{1-\chi}\ \right)}cos(x'\chi)K_0\left(\sqrt{y'^2+z'^2}\ \chi\right)d\chi + \int_1^{+\infty}\frac{\left(\varepsilon_0+\frac{A}{\chi^2}\left(2-\sqrt{1+\chi}\ \right)\right)cos(x'\chi)-\frac{A}{\chi^2}\sqrt{\chi-1}\ sin(x'\chi)}{\left(\varepsilon_0+\frac{A}{\chi^2}\left(2-\sqrt{1+\chi}\ \right)\right)^2+\frac{A^2}{\chi^4}(\chi-1)}K_0\left(\sqrt{y'^2+z'^2}\chi\right)d\chi\right\} \quad (14b)$$

$$B = \frac{q}{\pi\, v}\frac{E_{gap}}{\hbar} \quad (14c)$$

Let us study the behavior of (14b) and its integrand via plotting graphs of it:

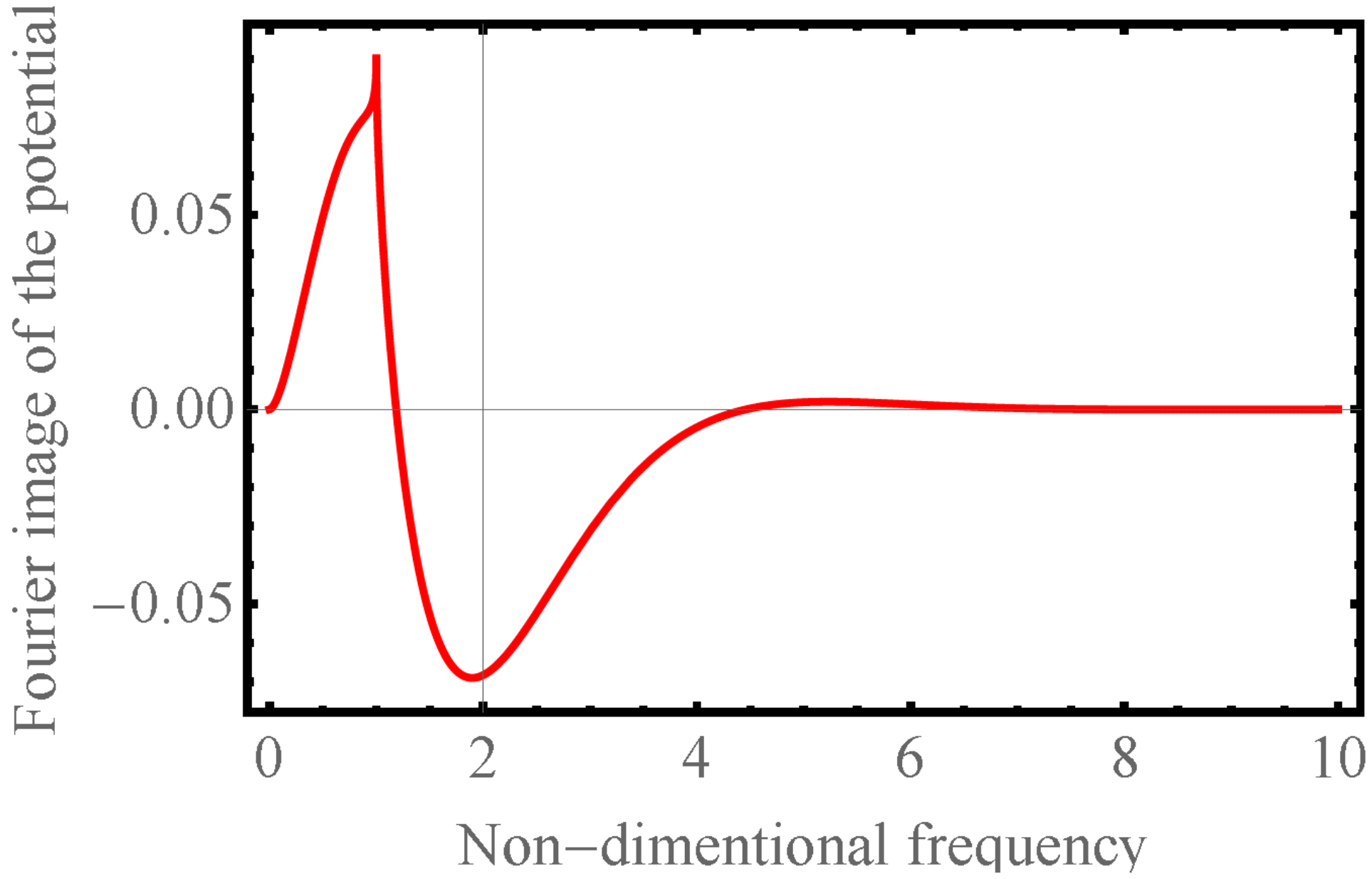

Fig. 1. The dependence of the Fourier image of the potential from the non-dimensional frequency $\chi$

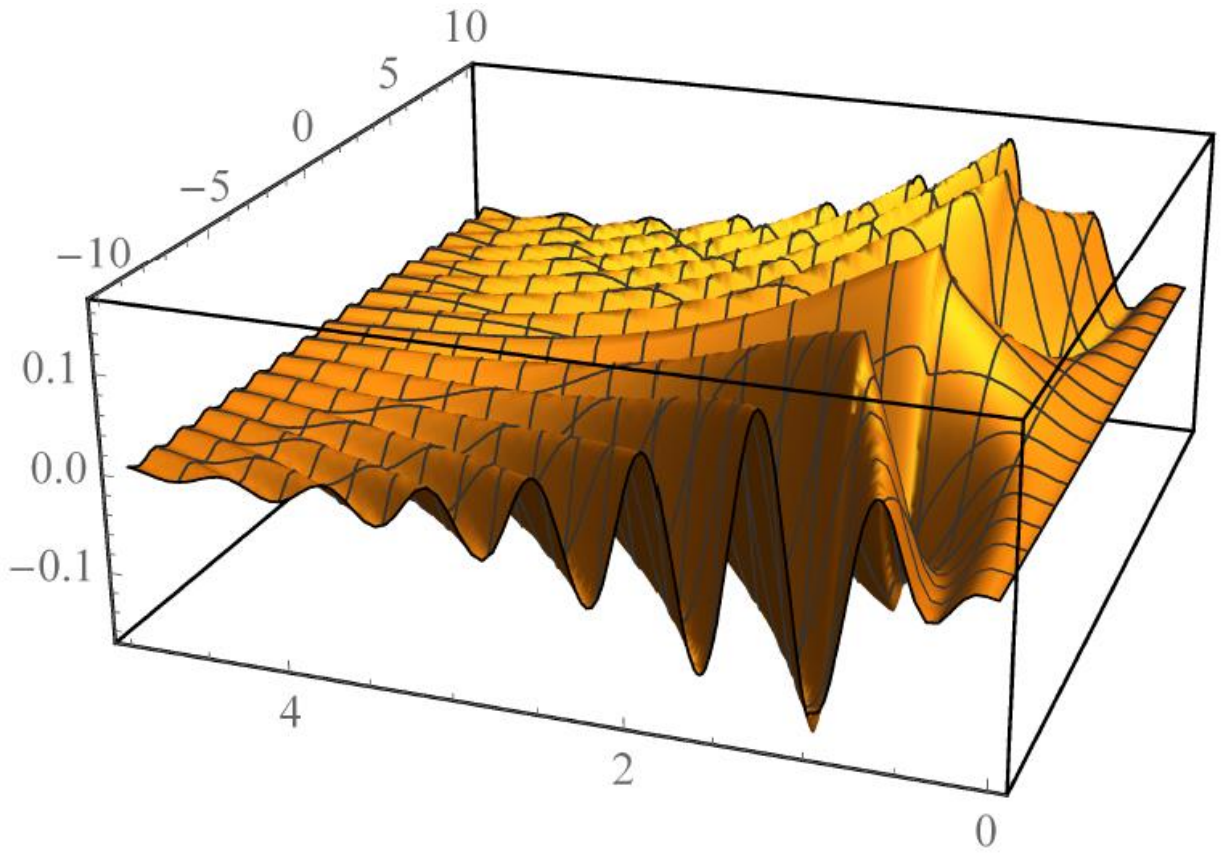

The image above shows the 3D plot of the Fourier Image of the electrostatic potential in the media. On the horizontal axis the non-dimensional frequency (longitudal) and the non-dimensional coordinate $x'$ are shown.

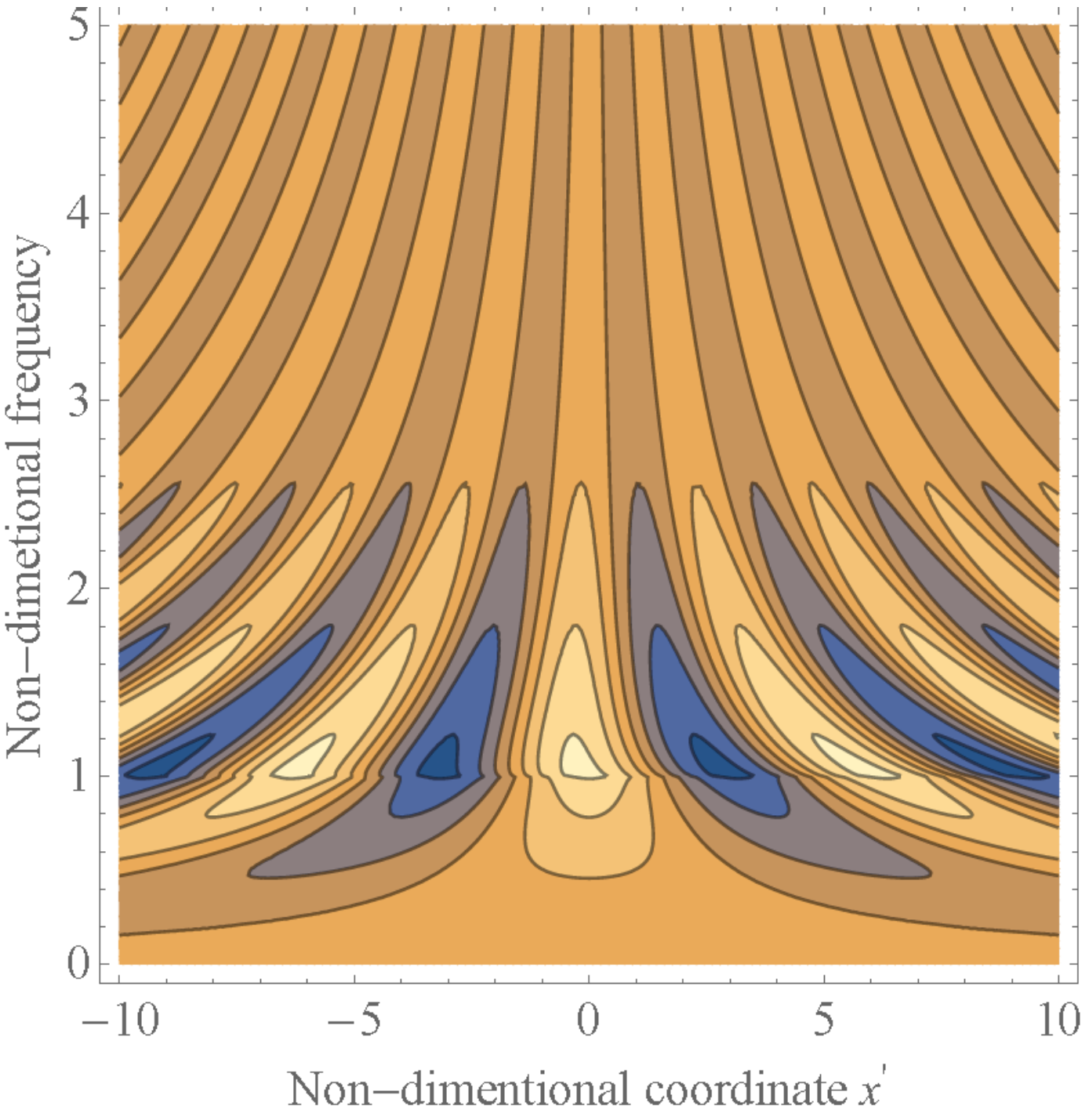

Fig. 2. The 3D plot of the Fourier image and the contour map of it

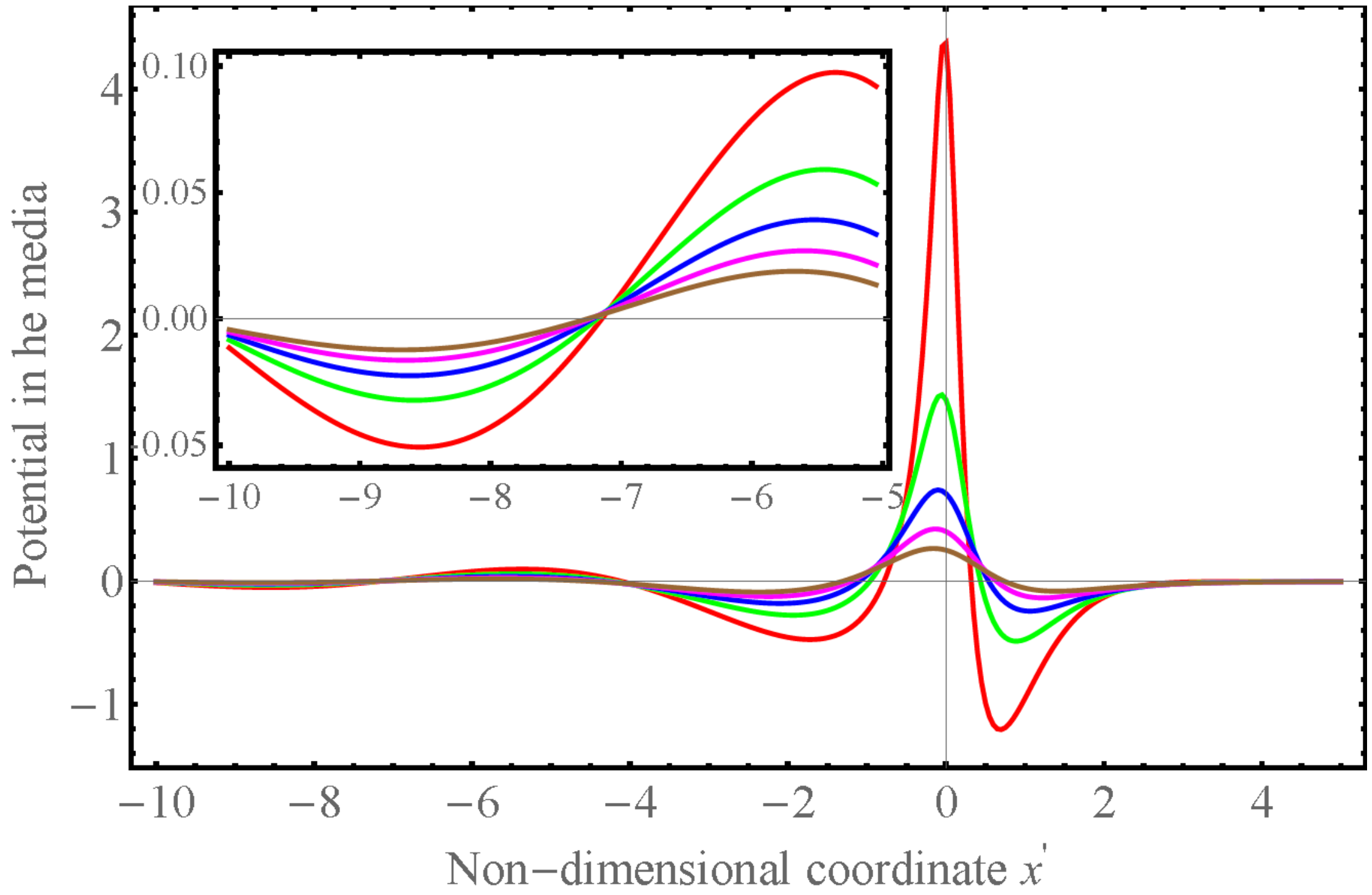

Fig. 3. The view of the dependence of the potential on the coordinate $x'$

Let us find the electrical field:

$$\vec{E} = -\vec{\nabla}\varphi \tag{15}$$

So, let us imply (15) considering (14b):

$$E_x = \frac{q}{\pi v}\frac{E_{gap}}{\hbar l}\left\{\int_0^1 \frac{1}{\varepsilon_0+\frac{A}{\chi^2}\left(2-\sqrt{1+\chi}+\sqrt{1-\chi}\right)}\chi \, sin(x'\chi)K_0\left(\sqrt{y'^2+z'^2}\,\chi\right)d\chi + \right.$$

$$\left.\int_1^{+\infty} \frac{\left(\varepsilon_0+\frac{A}{\chi^2}\left(2-\sqrt{1+\chi}\right)\right)sin(x'\chi)+\frac{A}{\chi^2}\sqrt{\chi-1}\,cos(x'\chi)}{\left(\varepsilon_0+\frac{A}{\chi^2}\left(2-\sqrt{1+\chi}\right)\right)^2+\frac{A^2}{\chi^4}(\chi-1)}\chi\, K_0\left(\sqrt{y'^2+z'^2}\chi\right)d\chi\right\} \tag{16a}$$

Also

$$\begin{cases} \frac{\partial}{\partial y}K_0\left(\sqrt{y'^2+z'^2}\,a\right) = -\frac{a\,y'}{l\sqrt{y'^2+z'^2}}K_1\left(\sqrt{y'^2+z'^2}\,a\right) \\ \frac{\partial}{\partial z}K_0\left(\sqrt{y'^2+z'^2}\,a\right) = -\frac{a\,z'}{l\sqrt{y'^2+z'^2}}K_1\left(\sqrt{y'^2+z'^2}\,a\right) \end{cases} \tag{16b}$$

Considering (16b) one can write the following expression:

$$\vec{E}_\perp = \vec{j}E_y + \vec{k}E_z =$$

$$\frac{q}{2\,\pi\,v}\frac{E_{gap}}{\hbar\,l}\frac{\vec{j}y'+\vec{k}z'}{\sqrt{{y'}^2+{z'}^2}}\left[\int_0^1 \frac{1}{\varepsilon_0+\frac{A}{\chi^2}(2-\sqrt{1+\chi}+\sqrt{1-\chi}\,)} cos(x'\chi)\,\chi K_1\left(\sqrt{{y'}^2+{z'}^2}\,\chi\right)d\chi + \right.$$

$$\left.\int_1^{+\infty} \frac{\left(\varepsilon_0+\frac{A}{\chi^2}(2-\sqrt{1+\chi}\,)\right)cos(x'\chi)-\frac{A}{\chi^2}\sqrt{\chi-1}\,sin(x'\chi)}{\left(\varepsilon_0+\frac{A}{\chi^2}(2-\sqrt{1+\chi}\,)\right)^2+\frac{A^2}{\chi^4}(\chi-1)}\chi\,K_1\left(\sqrt{{y'}^2+{z'}^2}\,\chi\right)d\chi\right] \tag{16c}$$

The power of the energy losses in the medium can be find via the following formula:

$$Q = -\frac{1}{4\,\pi}\iiint_V \vec{E}\frac{\partial\vec{D}}{\partial t}dV \tag{17}$$

In (17) we can do a Fourier transform:

$$\vec{D}(\vec{r},\omega) = \varepsilon(\omega)\,\vec{E}(\vec{r},\omega) \tag{18}$$

Let us now try and find the time derivative of the induction $\vec{D}$.

$$D_x = \frac{q}{\pi\,v}\frac{E_{gap}}{\hbar\,l}\left\{\int_0^1\left(\varepsilon_0+\frac{A}{\chi^2}(2-\sqrt{1+\chi}+\right.\right.$$

$$\left.\sqrt{1-\chi}\,)\right)\frac{1}{\varepsilon_0+\frac{A}{\chi^2}(2-\sqrt{1+\chi}+\sqrt{1-\chi}\,)}\,\chi\,sin(x'\chi)K_0\left(\sqrt{{y'}^2+{z'}^2}\,\chi\right)d\chi + \int_1^{+\infty}\left(\varepsilon_0+\right.$$

$$\left.\left.\frac{A}{\chi^2}(2-\sqrt{1+\chi}\,)\right)\frac{\left(\varepsilon_0+\frac{A}{\chi^2}(2-\sqrt{1+\chi}\,)\right)sin(x'\chi)+\frac{A}{\chi^2}\sqrt{\chi-1}\,cos(x'\chi)}{\left(\varepsilon_0+\frac{A}{\chi^2}(2-\sqrt{1+\chi}\,)\right)^2+\frac{A^2}{\chi^4}(\chi-1)}\,\chi\,K_0\left(\sqrt{{y'}^2+{z'}^2}\chi\right)d\chi\right\}$$

(19a)

$$\vec{D}_\perp = \vec{j}D_y + \vec{k}D_z = \frac{q}{2\,\pi\,v}\frac{E_{gap}}{\hbar\,l}\frac{\vec{j}y'+\vec{k}z'}{\sqrt{{y'}^2+{z'}^2}}\left[\int_0^1\left(\varepsilon_0+\frac{A}{\chi^2}(2-\sqrt{1+\chi}+\right.\right.$$

$$\left.\sqrt{1-\chi}\,)\right)\frac{1}{\varepsilon_0+\frac{A}{\chi^2}(2-\sqrt{1+\chi}+\sqrt{1-\chi}\,)}cos(x'\chi)\,\chi K_1\left(\sqrt{{y'}^2+{z'}^2}\,\chi\right)d\chi + \int_1^{+\infty}\left(\varepsilon_0+\right.$$

$$\left.\left.\frac{A}{\chi^2}(2-\sqrt{1+\chi}\,)\right)\frac{\left(\varepsilon_0+\frac{A}{\chi^2}(2-\sqrt{1+\chi}\,)\right)cos(x'\chi)-\frac{A}{\chi^2}\sqrt{\chi-1}\,sin(x'\chi)}{\left(\varepsilon_0+\frac{A}{\chi^2}(2-\sqrt{1+\chi}\,)\right)^2+\frac{A^2}{\chi^4}(\chi-1)}\chi\,K_1\left(\sqrt{{y'}^2+{z'}^2}\,\chi\right)d\chi\right]$$

(19b)

Let us find the physical value, which can clearly characterize the loss of the energy in the media:

$$\Delta x\iint_{-\infty}^{+\infty}\rho_Q dydz \tag{20}$$

The formula above is the loss of energy in an infinite layer of thickness $\Delta x$ in a unit time. $\rho_Q$ is the same as in formula (17). In order to find the energy, one needs to integrate (20) by time on during one period of oscillations. Let us try and look in this integration closer:

$$E = \Delta x\int_{-\infty}^{+\infty}dt\iint_{-\infty}^{+\infty}\rho_Q dydz \tag{21}$$

Let us consider the general form of (21) in a Fourier integral form:

$$\rho_E = \int_{-\infty}^{\infty} dt \frac{1}{(2\pi)^2} \iint_{-\infty}^{+\infty} \rho_Q(\omega_1, \omega_2) d\omega_1 d\,\omega_2 = -\frac{1}{4\,\pi} \int_{-\infty}^{+\infty} dt \frac{1}{(2\pi)^2} \iint_{-\infty}^{+\infty} \frac{1}{2} \Big[ i\omega_2 \vec{E}_\omega(\omega_1) \vec{D}_\omega^{\,*}(\omega_2) e^{-i(\omega_1-\omega_2)t} - i\omega_2 \vec{E}_\omega^{\,*}(\omega_1) \vec{D}_\omega(\omega_2) e^{-i(\omega_2-\omega_1)t} \Big] d\omega_1 d\,\omega_2 = -\frac{1}{8\,\pi^2} \iint_{-\infty}^{+\infty} \frac{1}{2} \Big[ i\omega_2 \vec{E}_\omega(\omega_1) \vec{D}_\omega^{\,*}(\omega_2) - i\omega_2 \vec{E}_\omega^{\,*}(\omega_1) \vec{D}_\omega(\omega_2) \Big] \delta(\omega_2 - \omega_1) d\omega_1 d\,\omega_2 = -\frac{1}{16\,\pi^2} \int_{-\infty}^{+\infty} \Big[ i\omega \vec{E}_\omega(\omega) \vec{D}_\omega^{\,*}(\omega) - i\omega_2 \vec{E}_\omega^{\,*}(\omega) \vec{D}_\omega(\omega) \Big] d\omega \quad (22)$$

Taking into account the view of the dielectric permittivity, one can obtain that

$$\rho_E = -\frac{1}{16\,\pi^2} \int_{-\infty}^{+\infty} \Big[ i\omega \vec{E}_\omega(\omega)(\varepsilon' - i\,\varepsilon'') \vec{E}_\omega^{\,*}(\omega) - i\omega \vec{E}_\omega^{\,*}(\omega)(\varepsilon' + i\,\varepsilon'') \vec{E}_\omega(\omega) \Big] d\omega \quad (23)$$

Considering the properties (6b) one can write that

$$\rho_E = -\frac{1}{16\,\pi^2} \int_{-\infty}^{+\infty} \Big[ i\omega \big|\vec{E}_\omega(\omega)\big|^2 (-i\,\varepsilon'') - i\omega \big|\vec{E}_\omega(\omega)\big|^2 i\,\varepsilon'' \Big] d\omega = -\frac{1}{4\,\pi^2} \int_0^{+\infty} \omega \big|\vec{E}_\omega(\omega)\big|^2 \varepsilon'' d\omega \quad (24a)$$

In our case we have the following expressions:

$$\varepsilon'' = \frac{A}{\chi^2} \sqrt{\chi - 1}\ \theta(\chi)\ \theta(\chi - 1) - \frac{A}{\chi^2} \sqrt{-\chi - 1}\ \theta(-\chi)\ \theta(-\chi - 1) \quad (25a)$$

$$\vec{E}_\omega(\omega) = -\vec{\nabla} \phi_\omega(\omega) \quad (25b)$$

And

$$\phi_\omega(\omega, \vec{r}) = \frac{q}{v\,\varepsilon(\omega)} e^{i\omega\frac{x}{v}} K_0\left(r_\perp \frac{|\omega|}{v}\right) \quad (25c)$$

So, we can find that

$$\vec{E}_\omega(\omega) = \frac{q}{v\,\varepsilon(\omega)} e^{i\omega\frac{x}{v}} \left[ i\omega \frac{\vec{e}_x}{v} K_0\left(r_\perp \frac{\omega}{v}\right) - \frac{\vec{r}_\perp}{r_\perp} \frac{\omega}{v} K_1\left(r_\perp \frac{\omega}{v}\right) \right] \quad (26)$$

The squared module of (26) can be found as follows:

$$\big|\vec{E}_\omega(\omega)\big|^2 = \frac{q^2}{v^2\,|\varepsilon(\omega)|^2} \frac{\omega^2}{v^2} \left[ {K_0}^2\left(r_\perp \frac{\omega}{v}\right) + {K_1}^2\left(r_\perp \frac{\omega}{v}\right) \right] \quad (27a)$$

If one substitutes (27a) into (24d) he finds that

$$\rho_E = -\frac{1}{4\,\pi^2} \frac{q^2}{v^4} \int_0^{+\infty} \frac{\omega^3 \varepsilon''(\omega)}{|\varepsilon(\omega)|^2} \left[ {K_0}^2\left(r_\perp \frac{\omega}{v}\right) + {K_1}^2\left(r_\perp \frac{\omega}{v}\right) \right] d\omega \quad (28)$$

Let us transform the variables in (28):

$$\chi = \frac{\hbar\,\omega}{E_{gap}} \quad (29a)$$

$$r_\perp{}' = \frac{r_\perp}{l} \quad (29b)$$

$$l = \frac{v}{\omega_{gap}}, \qquad \omega_{gap} = \frac{E_{gap}}{\hbar} \quad (29c)$$

Using (29a-c), we can obtain that

$$\rho_E = -\frac{1}{4\,\pi^2} \frac{q^2}{v^4\,{\omega_{gap}}^4} \int_0^{+\infty} \frac{\chi^3 \varepsilon''(\chi)}{|\varepsilon(\chi)|^2} \left[ {K_0}^2(r_\perp{}'\,\chi) + {K_1}^2(r_\perp{}'\,\chi) \right] d\chi \quad (30)$$

Here

$$\varepsilon' = \varepsilon_0 + \frac{A}{\chi^2}\left(2 - \sqrt{1+\chi}\,\theta(\chi) - \sqrt{1-\chi}\,\theta(-\chi) + \sqrt{1-|\chi|}\;\theta(1-\chi^2)\right)$$
$$\varepsilon'' = \frac{A}{\chi^2}\sqrt{\chi-1}\;\theta(\chi)\,\theta(\chi-1) - \frac{A}{\chi^2}\sqrt{-\chi-1}\,\theta(-\chi)\,\theta(-\chi-1) \qquad (31a)$$

$$|\varepsilon(\chi)|^2 = \varepsilon'^2 + \varepsilon''^2 \qquad (31b)$$

For example, for *InSt* we have the following parameters:

$$\varepsilon_0 = 16 \qquad (32a)$$

$$A = 2.6 \qquad (32b)$$

$$E_{gap} = 0.17\ eV \qquad (32c)$$

And the lattice constant

$$a = 6.5\ \text{Å} \qquad (33)$$

From Eq(32a) and Eq(32b) we can safely conclude that the value of real part of the permittivity never reaches zero. Now let us compare our "unit of length" $l = \frac{v}{\omega_{gap}}$ with the lattice constant, and derive from this comparison some condition for the speed of the particle. For our calculations to be reasonable we need that

$$l \gg a, \text{ so } v \gg a\frac{E_{gap}}{\hbar} = 6.5 \cdot 10^{-10}\ m\ \frac{0.17 \cdot 1.6 \cdot 10^{-19}\ J}{1.055 \cdot 10^{-34} J \cdot s} = \mathbf{1.67 \cdot 10^5 m/s} \qquad (34)$$

Now let us study the behavior of the integrated expression in the energy density (30) for different values of the radius $r_\perp'$. The expression can be written as follows:

$$\frac{\chi^3 \varepsilon''(\chi)}{|\varepsilon(\chi)|^2}\left[K_0{}^2(r_\perp'\,\chi) + K_1{}^2(r_\perp'\,\chi)\right] \qquad (35)$$

In (35) $\varepsilon''(\chi) \neq 0$ for $\chi > 1$. So, we can re-write it as follows:

$$\frac{A\,\chi^5\,\sqrt{\chi-1}}{A^2(\chi-1) + \left(\varepsilon_0\,\chi^2 + A\left(2-\sqrt{1+\chi}\right)\right)^2}\left[K_0{}^2(r_\perp'\,\chi) + K_1{}^2(r_\perp'\,\chi)\right] \qquad (36)$$

Or we can write

$$\frac{\chi^5\,\sqrt{\chi-1}}{\kappa^2\chi^4 + 2\,\kappa\,\chi^2\left(2-\sqrt{1+\chi}\right) + (2\chi+4) - 4\sqrt{1+\chi}}\,\frac{1}{A}\left[K_0{}^2(r_\perp'\,\chi) + K_1{}^2(r_\perp'\,\chi)\right] \qquad (37)$$

$$\text{Where } \kappa = \frac{\varepsilon_0}{A} = 6.153 \qquad (38)$$

Let us integrate (37) on the variable $r_\perp'$:

$$-\frac{1}{4\,\pi^2}\frac{q^2}{v^4\,\omega_{gap}{}^4}\frac{\chi^5\,\sqrt{\chi-1}}{\kappa^2\chi^4 + 2\,\kappa\,\chi^2(2-\sqrt{1+\chi}) + (2\chi+4) - 4\sqrt{1+\chi}}\,\frac{2\pi}{A}\int_a^\infty r_\perp'\left[K_0{}^2(r_\perp'\,\chi) + K_1{}^2(r_\perp'\,\chi)\right]dr_\perp' \qquad (39)$$

Let us evaluate the indefinite integral in Eq. (39). It is the sum of the two squared norms of the MacDonald functions. So, we can write that

$$\int r_\perp' K_0^{\,2}(r_\perp'\,\chi)dr_\perp' = \frac{(r_\perp')^2}{2}\left[K_0^{\,2}(r_\perp'\,\chi) - K_1^{\,2}(r_\perp'\,\chi)\right] \quad (40a)$$

$$\int r_\perp' K_1^{\,2}(r_\perp'\,\chi)dr_\perp' = \frac{(r_\perp')^2}{2}\left[K_1^{\,2}(r_\perp'\,\chi) - K_1(r_\perp'\,\chi)K_2(r_\perp'\,\chi)\right] \quad (40b)$$

Or the sum of Eqs(40a) and (40b) gives the following expression:

$$\frac{(r_\perp')^2}{2}\left[K_0^{\,2}(r_\perp'\,\chi) - K_1(r_\perp'\,\chi)K_2(r_\perp'\,\chi)\right]\Big|_a^\infty = -\frac{a^2}{2}\left[K_0^{\,2}(a\,\chi) - K_1(a\,\chi)K_2(a\,\chi)\right] \quad (41)$$

So, the energy of the losses in a layer of unit thickness can be written as is shown below. The layer is perpendicular to the movement of the particle.

$$\frac{\Delta E}{\Delta x} = -\frac{a^2}{4\pi\,A}\frac{q^2}{v^4\,\omega_{gap}{}^4}\int_1^\infty \frac{\chi^5\,\sqrt{\chi-1}\left[K_1(a\,\chi)K_2(a\,\chi) - K_0{}^2(a\,\chi)\right]}{\kappa^2\chi^4 + 2\,\kappa\,\chi^2(2-\sqrt{1+\chi}) + (2\chi+4) - 4\sqrt{1+\chi}}\,d\chi \quad (42)$$

Let us try to write some analogue of the motion equation for our charged particle:

$$\frac{d}{dx}T = \frac{dE}{dx} \quad (43)$$

Where *T* is the kinetic energy of the particle. The right hand side holds the same value as Eq.(42). So, we can write as follows

$$m\,v\,v' = -\frac{a^2}{4\pi\,A}\frac{q^2}{v^4\,\omega_{gap}{}^4}I \quad (44)$$

In (44) *I* is the value of the integral. It does not depend on the speed of the particle. So, we can write the following expression:

$$-\frac{4\pi\,A\,m\,\omega_{gap}{}^4}{a^2q^2I}\,v^5\,dv = dx, v(0) = v_0 \quad (45)$$

Or after the integration we can write that

$$\frac{2\pi\,A\,m\,\omega_{gap}{}^4}{3\,a^2q^2I}(v_0{}^6 - v^6) = x, \text{ or } v = \left[v_0{}^6 - \frac{3\,a^2q^2I}{2\pi\,A\,m\,\omega_{gap}{}^4}x\right]^{\frac{1}{6}} \quad (46)$$

$$\rho_E = -\frac{1}{16\,\pi^3}\frac{q^2\,E_{gap}{}^2}{v\hbar^2\,l^2}\iint_{-\infty}^{+\infty} d\chi_1 d\chi_2 \int_{-\frac{x-vT}{2l}}^{\frac{x-vT}{2l}} \frac{\chi_1\,\chi_2\,\varepsilon'(\chi_2)\varepsilon'(\chi_1)\varepsilon''(\chi_2)}{|\varepsilon(\chi_2)|^2\,|\varepsilon(\chi_1)|^2} cos\left(x'(\chi_1 + \chi_2)\right)\left[K_0(r_\perp\,\chi_1)K_0(r_\perp\,\chi_2) - K_1(r_\perp\,\chi_1)K_1(r_\perp\,\chi_2)\right]dx' \quad (34c)$$

After we do the integration on the dimensionless coordinate in (34c), one can obtain that

$$\rho_E = -\frac{1}{8\,\pi^3}\frac{q^2\,E_{gap}{}^2}{v\hbar^2\,l^2}\iint_{-\infty}^{+\infty}\frac{\chi_1\,\chi_2\,\varepsilon'(\chi_2)\varepsilon'(\chi_1)\varepsilon''(\chi_2)}{(\chi_1+\chi_2)|\varepsilon(\chi_2)|^2\,|\varepsilon(\chi_1)|^2} sin\left(\frac{x-vT}{2l}(\chi_1 + \chi_2)\right)\left[K_0(r_\perp\,\chi_1)K_0(r_\perp\,\chi_2) - K_1(r_\perp\,\chi_1)K_1(r_\perp\,\chi_2)\right]d\chi_1 d\chi_2 \quad (35)$$

# Supplement 2

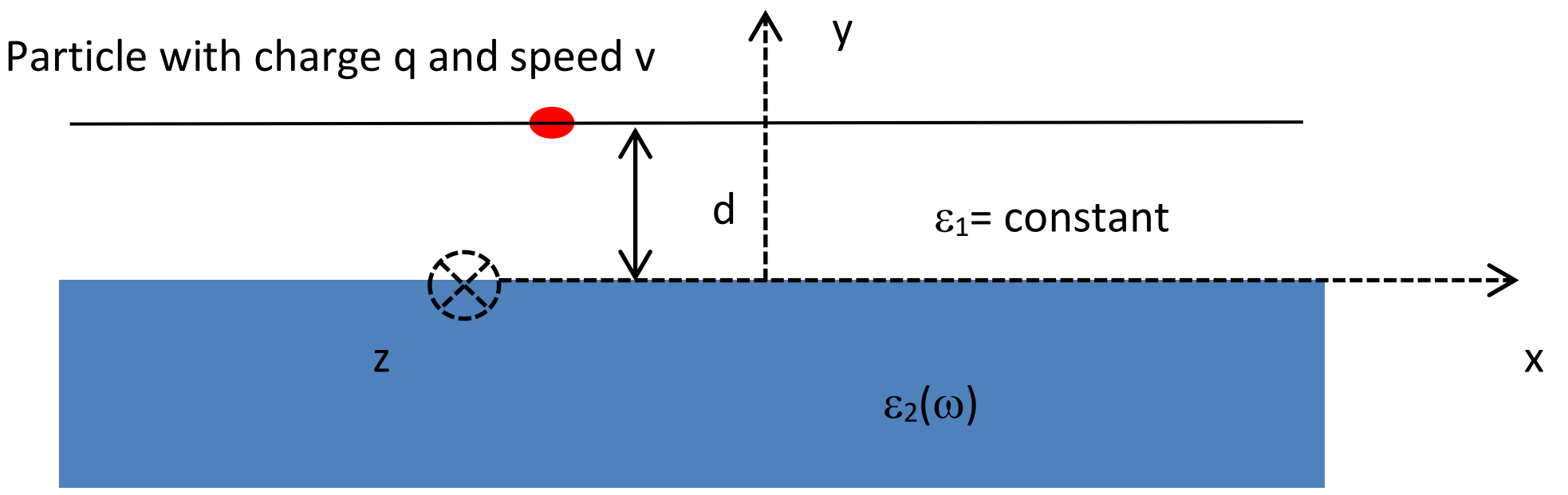

The equations can be written as follows:

$$\begin{cases} div(\vec{D}) = 4\,\pi\, q\, \delta(x - vt)\delta\,(y - d)\delta(z) \\ \vec{D}(x, y, z, t) = \int_{-\infty}^{0} \varepsilon(t - \tau)\vec{E}\,(x, y, z, \tau)d\tau \end{cases} \tag{1}$$

Boundary conditions on the induction field and the electric field:

$$\begin{cases} \left(D_y^{(1)} - D_y^{(2)}\right)\Big|_{y=0} = 0 \\ \left(E_x^{(1)} - E_x^{(2)}\right)\Big|_{y=0} = 0 \end{cases} \tag{2}$$

Let us do a Fourier transform on the variable *t*:

$$\vec{D}(x, y, z, \omega) = \ \varepsilon(\omega)\,\vec{E}(x, y, z, \omega) \tag{3a}$$

$$div\vec{D}(\,\omega) = \ \int_{-\infty}^{+\infty} 4\,\pi\, q\, \delta(x - vt)\delta\,(y - d)\delta(z)e^{i\omega t}dt = \frac{4\,\pi\, q}{v}\ e^{+i\omega\frac{x}{v}}\delta\,(y - d)\delta(z) \tag{3b}$$

Let use the scalar electrostatic potential, because $rot(\vec{E}) = 0$:

$$\vec{E}(x, y, z, \omega) = \ -\vec{\nabla}\varphi(x, y, z, \omega) \tag{4a}$$

$$div\left(\ \varepsilon(\omega)\ \cdot\left(-\vec{\nabla}\varphi(\omega)\right)\right) = 4\,\pi\, q\ e^{+i\omega\frac{x}{v}}\delta\,(y - d)\delta(z) \tag{4b}$$

So, finally we can obtain such equations and boundary conditions:

$$\begin{cases} \Delta\varphi^{(1)}(\omega, \vec{r}) = -\frac{4\,\pi\, q}{v\,\varepsilon_1}\ e^{+i\omega\frac{x}{v}}\delta\,(y - d)\delta(z), \qquad y > 0 \\ \varepsilon_2(\omega)\Delta\varphi^{(2)}(\omega, \vec{r}) = 0, \qquad y < 0 \\ \varphi^{(1)}\Big|_{y=0} = \varphi^{(2)}\Big|_{y=0} \\ \varepsilon_1 \frac{\partial\varphi^{(1)}}{\partial y}\Big|_{y=0} = \varepsilon_2 \frac{\partial\varphi^{(2)}}{\partial y}\Big|_{y=0} \end{cases} \tag{5}$$

1) *Let us solve system (5) directly, via the use of integral transformations:*

As one can easily see, the source in the first equation is a periodic function of the variable *x*. So, it is very natural to find the solution of the system above like a periodic function of *x:*

$$\varphi^{(n)}(x,y,z,\omega) = \psi_n(y,z,\omega)e^{+i\omega\frac{x}{v}} \qquad n=1,2 \tag{6}$$

Using this expression (6), our equations can be transformed like is shown below:

$$\begin{cases} \Delta\psi_1(y,z,\omega) - \frac{\omega^2}{v^2}\psi_1(y,z,\omega) = -\frac{4\,\pi\,q}{v\,\varepsilon_1}\,\delta\,(y-d)\delta(z), \qquad y>0 \\ \Delta\psi_2(y,z,\omega) - \frac{\omega^2}{v^2}\psi_2(y,z,\omega) = 0, \qquad y<0 \\ \psi_1|_{y=0} = \psi_2\big|_{y=0} \\ \varepsilon_1\frac{\partial\psi_1}{\partial y}\Big|_{y=0} = \varepsilon_2\frac{\partial\psi_2}{\partial y}\Big|_{y=0} \end{cases} \tag{7}$$

It is also important to note that the module of the potential must be limited, when the coordinates tend to infinity. Now, let us do a Fourier transform on the variable *z*:

$$\widetilde{\psi_n}(y,k,\omega) = \int_{-\infty}^{+\infty} e^{ikz}\psi_n(y,z,\omega)\,dz \tag{8}$$

$$\int_{-\infty}^{+\infty} e^{ikz}\Delta\psi_n(y,z,\omega)\,dz = \int_{-\infty}^{+\infty} e^{ikz}\left[\frac{\partial^2}{\partial y^2}\psi_n(y,z,\omega) + \frac{\partial^2}{\partial z^2}\psi_n(y,z,\omega)\right]dz = \frac{\partial^2}{\partial y^2}\widetilde{\psi_n}(y,k,\omega) + e^{ikz}\cdot\frac{\partial}{\partial z}\psi_n(y,z,\omega)\Big|_{-\infty}^{+\infty} - i\,k\int_{-\infty}^{+\infty} e^{ikz}\left[\frac{\partial}{\partial z}\psi_n(y,z,\omega)\right]dz \tag{8a}$$

The second term in the right hand side turns to zero due to the limited field. If one does integration by parts in the third term, he will obtain such expression:

$$\int_{-\infty}^{+\infty} e^{ikz}\Delta\psi_n(y,z,\omega)\,dz = \frac{\partial^2}{\partial y^2}\widetilde{\psi_n}(\omega,k,y) - k^2\widetilde{\psi_n}(\omega,k,y) \tag{8b}$$

So, using expression (8b) and doing a Fourier transformation of the delta function, we can finally obtain the equations and conditions:

$$\begin{cases} \frac{d^2}{dy^2}\widetilde{\psi_1}(\omega,k,y) - \left(k^2+\frac{\omega^2}{v^2}\right)\widetilde{\psi_1}(\omega,k,y) = -\frac{4\,\pi\,q}{v\,\varepsilon_1(\omega)}\,\delta\,(y-d), \qquad y>0, \\ \frac{d^2}{dy^2}\widetilde{\psi_2}(\omega,k,y) - \left(k^2+\frac{\omega^2}{v^2}\right)\widetilde{\psi_2}(\omega,k,y) = 0, \qquad y<0, \\ \widetilde{\psi_1}\big|_{y=0} = \widetilde{\psi_2}\big|_{y=0} \\ \varepsilon_1\frac{d\widetilde{\psi_1}}{dy}\Big|_{y=0} = \varepsilon_2\frac{d\widetilde{\psi_2}}{dy}\Big|_{y=0} \end{cases} \tag{9}$$

These equations are Helmholtz equations; the first of them is heterogeneous. The conditions are both homogeneous. They have hyperbolic function type solutions:

$$\widetilde{\psi_1}(\omega,k,y) = C_1e^{-y\sqrt{k^2+\frac{\omega^2}{v^2}}} + \widetilde{\psi_1}^{partial}(\omega,k,y) \tag{10a}$$

$$\widetilde{\psi_2}(\omega,k,y) = C_2e^{+y\sqrt{k^2+\frac{\omega^2}{v^2}}} \tag{10b}$$

Let us find the solution of the first equation in such form:

$$\widetilde{\psi_1}^{partial}(\omega,k,y) \; = \; A\, e^{-|y-d|\sqrt{k^2+\frac{\omega^2}{v^2}}} \tag{11}$$

By substitution of this formula to the first equation, we obtain such expression:

$$A\sqrt{k^2+\frac{\omega^2}{v^2}}\frac{d}{dy}\left(-sign(y-d)\;\; e^{-|y-d|\sqrt{k^2+\frac{\omega^2}{v^2}}}\right)-\left(k^2+\frac{\omega^2}{v^2}\right)A\, e^{-|y-d|\sqrt{k^2+\frac{\omega^2}{v^2}}}=-\frac{4\,\pi\, q}{v\,\varepsilon_1(\omega)}\,\delta\,(y-d) \tag{11a}$$

If we take the derivative second time, we obtain the following equation on A:

$$A\,\sqrt{k^2+\frac{\omega^2}{v^2}}\left(-2\,\delta(y-d)\;\; e^{-|y-d|\sqrt{k^2+\frac{\omega^2}{v^2}}}\right)=-\frac{4\,\pi\, q}{v\,\varepsilon_1(\omega)}\,\delta\,(y-d) \tag{11b}$$

Here we take into account, that$[sign(y-d)]^2\equiv 1$. Due to this property, the second term cancels with the first term in the derivative. Also it is important to note that the Dirac delta-function has the following property:

$$\delta(y-d)\,f(y)=\;\frac{f(d+0)+f(d-0)}{2}\delta(y-d) \tag{12}$$

Taking into account (12) and (11b), one can obtain such formula:

$$A\,\sqrt{k^2+\frac{\omega^2}{v^2}}=\frac{2\,\pi\, q}{v\,\varepsilon_1(\omega)} \tag{13}$$

Now, let us write down the general solution of the system (9):

$$\widetilde{\psi_1}(\omega,k,y)=C_1 e^{-y\sqrt{k^2+\frac{\omega^2}{v^2}}}+\frac{2\,\pi\, q}{v\,\varepsilon_1(\omega)}\frac{1}{\sqrt{k^2+\frac{\omega^2}{v^2}}}\; e^{-|y-d|\sqrt{k^2+\frac{\omega^2}{v^2}}}\;\;, \tag{14a}$$

$$\widetilde{\psi_2}(\omega,k,y)=C_2 e^{y\sqrt{k^2+\frac{\omega^2}{v^2}}}\;. \tag{14b}$$

Now we need to apply the boundary conditions. From the conditions of the first type we can obtain such formula:

$$C_1+\frac{2\,\pi\, q}{v\,\varepsilon_1(\omega)}\frac{1}{\sqrt{k^2+\frac{\omega^2}{v^2}}}\; e^{-d\sqrt{k^2+\frac{\omega^2}{v^2}}}=C_2 \tag{15a}$$

When applying the second one, such equation can be obtained:

$$\varepsilon_1(\omega)\cdot\sqrt{k^2+\frac{\omega^2}{v^2}}\cdot\left[-C_1+\frac{2\,\pi\, q}{v\,\varepsilon_1(\omega)}\frac{1}{\sqrt{k^2+\frac{\omega^2}{v^2}}}\; e^{-d\sqrt{k^2+\frac{\omega^2}{v^2}}}\right]=\varepsilon_2(\omega)\,\sqrt{k^2+\frac{\omega^2}{v^2}}\cdot C_2 \tag{15b}$$

Therefore, from (15a, b), we have obtained a system on $C_1$ and $C_2$:

$$\begin{cases} C_1 - C_2 = -\frac{2\,\pi\, q}{v\,\varepsilon_1(\omega)} \frac{1}{\sqrt{k^2+\frac{\omega^2}{v^2}}}\ e^{-d\sqrt{k^2+\frac{\omega^2}{v^2}}} \\ \varepsilon_1 \cdot C_1 + \varepsilon_2 \cdot C_2 = \frac{2\,\pi\, q}{v}\ \frac{1}{\sqrt{k^2+\frac{\omega^2}{v^2}}}\ e^{-d\sqrt{k^2+\frac{\omega^2}{v^2}}} \end{cases} \tag{16}$$

Using elementary linear algebra, one can find both constants:

$$\begin{cases} C_1 = -\frac{\varepsilon_1(\omega)-(\omega)}{\varepsilon_1(\omega)+\varepsilon_2(\omega)} \frac{2\pi\, q}{v\,\varepsilon_1(\omega)} \frac{1}{\sqrt{k^2+\frac{\omega^2}{v^2}}}\ e^{-d\sqrt{k^2+\frac{\omega^2}{v^2}}} \\ C_2 = \frac{4\,\pi\, q}{v\,(\varepsilon_1+\varepsilon_2)} \frac{1}{\sqrt{k^2+\frac{\omega^2}{v^2}}}\ e^{-d\sqrt{k^2+\frac{\omega^2}{v^2}}} \end{cases} \tag{17}$$

Now we can write down the expressions for $\widetilde{\psi_k}(\omega, k, y)$ :

$$\widetilde{\psi_1}(\omega, k, y) = \frac{2\,\pi\, q}{v\,\varepsilon_1(\omega)\sqrt{\left(k^2+\frac{\omega^2}{v^2}\right)}}\left[-\frac{\varepsilon_2(\omega)-\varepsilon_1(\omega)}{\varepsilon_1(\omega)+\varepsilon_2(\omega)}\ e^{-(d+y)\sqrt{k^2+\frac{\omega^2}{v^2}}} +\ e^{-|y-d|\sqrt{k^2+\frac{\omega^2}{v^2}}}\right],\quad y>0 \tag{18a}$$

$$\widetilde{\psi_2}(\omega, k, y) = \frac{4\,\pi\, q}{v\,(\varepsilon_1(\omega)+\varepsilon_2(\omega))}\ \frac{1}{\sqrt{k^2+\frac{\omega^2}{v^2}}}\ e^{(y-d)\sqrt{k^2+\frac{\omega^2}{v^2}}} \qquad y<0 \tag{18b}$$

Now we can do the reverse Fourier transformation two times (on ω and on $k$) and obtain the expressions for the electrostatic potentials $\varphi_1$ and $\varphi_2$. Also we must use the following formulas for the dielectric permittivitys (Anselm, page 172):

$$\begin{gathered} \varepsilon_1 = \varepsilon = constant \\ \varepsilon_2 = \chi_\infty \frac{\omega_{lo}^2-\omega^2-2i\gamma\omega}{\omega_{to}^2-\omega^2-2i\gamma\omega}\ \ \ldots \end{gathered} \tag{19}$$

In (19) $N$ is the concentration of charge carriers, $e$ is the elementary charge, and $m$ is the electron mass. Now let us write down the formulas for the reverse Fourier transform, taking into account formula (6):

$$\psi_n(\omega, z, y) = \frac{1}{2\,\pi}\int_{-\infty}^{+\infty} e^{-ikz}\,\widetilde{\psi_n}(\omega, k, y)dk \tag{20a}$$

$$\varphi^{(n)}(\vec{r}, t) = \frac{1}{2\,\pi}\int_{-\infty}^{+\infty} e^{-i\omega t}\,\varphi^{(n)}(\omega)d\omega = \left(\frac{1}{2\,\pi}\right)^2 \int_{-\infty}^{+\infty} d\omega\ e^{-i\omega(t-\frac{x}{v})} \int_{-\infty}^{+\infty} e^{-ikz}\,\widetilde{\psi_n}(\omega, k, y)dk \tag{20b}$$

Some useful integrals could be found in [Градштейн && Рыжик, 4-е издание, page 512, equation 3.961-2]:

$$\int_0^\infty \frac{cos(k\,\beth)}{\sqrt{k^2+\vartheta^2}}\ e^{-\aleph\sqrt{k^2+\vartheta^2}}dk = K_0\left(\vartheta\,\sqrt{\beth^2+\aleph^2}\right) \tag{21}$$

Let us transform the integrals (20a) to the table view (21):

$$\psi_1(\omega,z,y)=\frac{1}{2\,\pi}\frac{2\,\pi\,q}{v\,\varepsilon_1(\omega)}\int_{-\infty}^{+\infty}\frac{1}{\sqrt{\left(k^2+\frac{\omega^2}{v^2}\right)}}e^{-ikz}\left[-\frac{\varepsilon_2(\omega)-\varepsilon_1(\omega)}{\varepsilon_1(\omega)+\varepsilon_2(\omega)}\;e^{-(d+y)\sqrt{k^2+\frac{\omega^2}{v^2}}}+\;e^{-|y-d|\sqrt{k^2+\frac{\omega^2}{v^2}}}\right]dk=$$

$$A(\omega)\int_{-\infty}^{+\infty}\frac{1}{\sqrt{\left(k^2+\frac{\omega^2}{v^2}\right)}}(\cos(kz)+i\sin(kz))\left[-\frac{\varepsilon_2(\omega)-\varepsilon_1(\omega)}{\varepsilon_1(\omega)+\varepsilon_2(\omega)}\;e^{-(d+y)\sqrt{k^2+\frac{\omega^2}{v^2}}}+\;e^{-|y-d|\sqrt{k^2+\frac{\omega^2}{v^2}}}\right]dk\ (22)$$

If we take into account that sin $(kz)$ is an uneven function and it is multiplied by an an even function of k, we can rewrite the expression (22) in such form:

$$\psi_1(\omega,z,y)=2A(\omega)\int_{0}^{+\infty}\frac{1}{\sqrt{\left(k^2+\frac{\omega^2}{v^2}\right)}}\cos(kz)\left[-\frac{\varepsilon_2(\omega)-\varepsilon_1(\omega)}{\varepsilon_1(\omega)+\varepsilon_2(\omega)}\;e^{-(d+y)\sqrt{k^2+\frac{\omega^2}{v^2}}}+\right.$$

$$\left.e^{-|y-d|\sqrt{k^2+\frac{\omega^2}{v^2}}}\right]dk\ (23)$$

Where $A(\omega)=\frac{q}{v\,\varepsilon_1(\omega)}$ (23a)

Now we can directly use the formula (21):

$$\psi_1(\omega,z,y)=2A(\omega)\int_{0}^{\infty}\frac{cos(k\,ב)}{\sqrt{k^2+\vartheta^2}}\left[B(\omega)e^{-\aleph_1\sqrt{k^2+\vartheta^2}}+\;e^{-\aleph_2\sqrt{k^2+\vartheta^2}}\right]dk=$$

$$2A(\omega)\left[B(\omega)K_0\left(\vartheta\sqrt{ב^2+{\aleph_1}^2}\right)+K_0\left(\vartheta\sqrt{ב^2+{\aleph_2}^2}\right)\right]\quad(24)$$

$$ב=z$$
$$\vartheta=\frac{\omega}{v}$$
$$\aleph_1=(d+y)\quad(24a)$$
$$\aleph_1=|y-d|$$
$$B(\omega)=-\frac{\varepsilon_2(\omega)-\varepsilon_1(\omega)}{\varepsilon_1(\omega)+\varepsilon_2(\omega)}$$

Finally, we have obtained such expression:

$$\psi_1(\omega,z,y)=2\frac{q}{v\,\varepsilon_1(\omega)}\left[-\frac{\varepsilon_2(\omega)-\varepsilon_1(\omega)}{\varepsilon_1(\omega)+\varepsilon_2(\omega)}K_0\left(\frac{|\omega|}{v}\sqrt{z^2+(y+d)^2}\right)+K_0\left(\frac{|\omega|}{v}\sqrt{z^2+|y-d|^2}\right)\right]$$
(24b)

Analogically, we can do the same transformations to evaluate $\psi_2(\omega,z,y)$:

$$\psi_2(\omega,z,y)=\frac{2\,q}{v\,(\varepsilon_1(\omega)+\varepsilon_2(\omega))}\int_{-\infty}^{+\infty}e^{-ikz}\;\frac{1}{\sqrt{k^2+\frac{\omega^2}{v^2}}}\;e^{(y-d)\sqrt{k^2+\frac{\omega^2}{v^2}}}dk=$$

$$\frac{4\,q}{v\,(\varepsilon_1(\omega)+\varepsilon_2(\omega))}\int_{0}^{+\infty}\cos(kz)\;\frac{1}{\sqrt{k^2+\frac{\omega^2}{v^2}}}\;e^{-(d-y)\sqrt{k^2+\frac{\omega^2}{v^2}}}dk=\frac{4\,q}{v\,(\varepsilon_1(\omega)+\varepsilon_2(\omega))}K_0\left(\frac{|\omega|}{v}\sqrt{z^2+(d-y)^2}\right)$$

(24c)

Now we can use the formula (6): $\varphi^{(n)}(x,y,z,\omega)=\psi_n(y,z,\omega)e^{+i\omega\frac{x}{v}}$

So, we have obtained the frequency spectrum of the electrostatic potential:

$$\begin{cases} \varphi^{(1)}(x,y,z,\omega) = 2\frac{q}{v\,\varepsilon}\left[K_0\left(\frac{|\omega|}{v}\sqrt{z^2+(y-d)^2}\right) + \frac{\varepsilon-\varepsilon_2(\omega)}{\varepsilon+\varepsilon_2(\omega)}K_0\left(\frac{|\omega|}{v}\sqrt{z^2+(y+d)^2}\right)\right]e^{i\omega\frac{x}{v}}, y>0 \\ \varphi^{(2)}(x,y,z,\omega) = \frac{4\,q}{v\,(\varepsilon+\varepsilon_2(\omega))}\,e^{i\omega\frac{x}{v}}K_0\left(\frac{|\omega|}{v}\sqrt{z^2+(d-y)^2}\right), y<0 \end{cases}$$

(25)

In (25) $K_0$ is the modified Bessel function of zero order. In some sources it is often called "MacDonald function". Also $\varepsilon_2 = \chi_\infty \frac{\omega_{lo}^2-\omega^2-2i\gamma\omega}{\omega_{to}^2-\omega^2-2i\gamma\omega}$.

*Now let us solve the system (5) using Green`s function:*

$$\begin{cases} \Delta G^{(1)}(\omega,\vec{r}) = -\frac{4\pi q}{v\,\varepsilon_1}\,\delta(x-\tilde{x})\delta(y-\tilde{y})\delta(z-\tilde{z}), \quad y>0 \\ \varepsilon_2(\omega)\Delta G^{(2)}(\omega,\vec{r}) = 0, \quad y<0 \\ G^{(1)}\big|_{y=0} = G^{(2)}\big|_{y=0} \\ \varepsilon_1\frac{\partial G^{(1)}}{\partial y}\Big|_{y=0} = \varepsilon_2\frac{\partial G^{(2)}}{\partial y}\Big|_{y=0} \end{cases} \qquad (26)$$

When we solve (25), we can apply the well-known formula (26):

$$\varphi = \iiint_{-\infty}^{+\infty} G(x,\tilde{x},y,\tilde{y},z,\tilde{z})f(\tilde{x},\tilde{y},\tilde{z})d\tilde{x}\,d\tilde{y}\,d\tilde{z} \qquad (26)$$

Where $f(\tilde{x},\tilde{y},\tilde{z}) = \frac{1}{v}e^{+i\omega\frac{\tilde{x}}{v}}\delta(\tilde{y}-d)\delta(\tilde{z})$ is the source in system (5).

The problem of finding the Green`s function is a well-known problem about a charge carrier above an interface between two dielectrical half-spaces. It can be solved with the mirror image method:

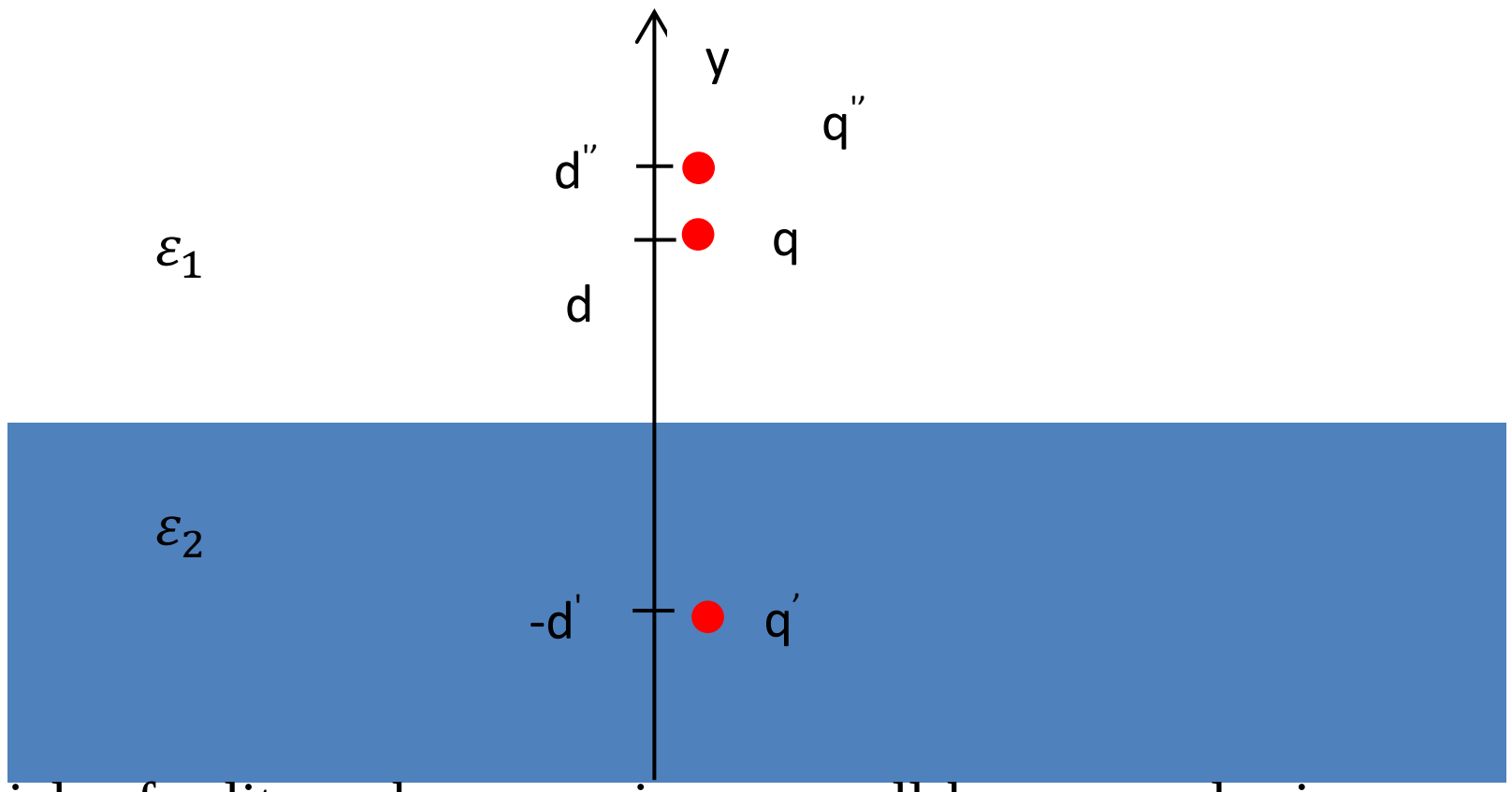

The potentials of solitary charge carriers are well-known, and using superposition method one can write the formulas below:

$$G^{(1)} = \frac{q}{\varepsilon_1\sqrt{x^2+(y-d)^2+z^2}} + \frac{q'}{\varepsilon_1\sqrt{x^2+(y+d')^2+z^2}} \qquad (27a)$$

$$G^{(2)} = \frac{q}{\varepsilon_2\sqrt{x^2+(y-d)^2+z^2}} + \frac{q''}{\varepsilon_2\sqrt{x^2+(y-d'')^2+z^2}} \qquad (27b)$$

The mirror image charges and the distances can be found from the boundary conditions:

$$\begin{cases} \dfrac{q}{\varepsilon_1\sqrt{x^2+d^2+z^2}} + \dfrac{q'}{\varepsilon_1\sqrt{x^2+d'^2+z^2}} = \dfrac{q''}{\varepsilon_2\sqrt{x^2+d''^2+z^2}} + \dfrac{q}{\varepsilon_2\sqrt{x^2+d^2+z^2}} \\ \dfrac{q\,d}{\sqrt{(x^2+d^2+z^2)^3}} - \dfrac{q'd'}{\sqrt{\left(x^2+d'^2+z^2\right)^3}} = \dfrac{q''d''}{\sqrt{\left(x^2+d''^2+z^2\right)^3}} + \dfrac{q\,d}{\sqrt{(x^2+d^2+z^2)^3}} \end{cases}$$

After simple algebraic transformation, we obtain such equations:

$$\begin{cases} \frac{q}{\varepsilon_1\sqrt{(x^2+d^2+z^2}} + \frac{q'}{\varepsilon_1\sqrt{x^2+d'^2+z^2}} = \frac{q''}{\varepsilon_2\sqrt{x^2+d''^2+z^2}} + \frac{q}{\varepsilon_2\sqrt{x^2+d^2+z^2}} \\ \frac{-q''d''}{\sqrt{\left(x^2+d''^2+z^2\right)^3}} = \frac{q'd'}{\sqrt{\left(x^2+d'^2+z^2\right)^3}} \end{cases} \tag{28}$$

From the second equation in system (28) we can obtain that $-q''d'' = q'd'$ and that $d'' = d'$ .

So, $q'' = -q'$. By substitution of this expression to the first equation, one can obtain, that

$$\frac{q}{\varepsilon_1\sqrt{(x^2+d^2+z^2}} + \frac{q'}{\varepsilon_1\sqrt{x^2+d'^2+z^2}} = \frac{-q'}{\varepsilon_2\sqrt{x^2+d'^2+z^2}} + \frac{q}{\varepsilon_2\sqrt{x^2+d^2+z^2}}$$

$$q'\left(\frac{1}{\varepsilon_1}+\frac{1}{\varepsilon_2}\right)\frac{1}{\sqrt{x^2+d'^2+z^2}} = q\left(\frac{1}{\varepsilon_2}-\frac{1}{\varepsilon_1}\right)\frac{1}{\sqrt{x^2+d^2+z^2}} \tag{28a}$$

From (28a) it is obvious, that $d = d'$ and $q' = q\left(\frac{1}{\varepsilon_2}-\frac{1}{\varepsilon_1}\right)\frac{\varepsilon_1\,\varepsilon_2}{\varepsilon_1+\varepsilon_2} = q\frac{\varepsilon_1-\varepsilon_2}{\varepsilon_1+\varepsilon_2}$. Finally, we have obtained such expressions for the mirror image charges and distances:

$$\begin{cases} d = d' = d'' \\ q' = q\frac{\varepsilon_1-\varepsilon_2}{\varepsilon_1+\varepsilon_2} \\ q'' = q\frac{\varepsilon_2-\varepsilon_1}{\varepsilon_1+\varepsilon_2} \end{cases} \tag{29}$$

Now, let us substitute (29) into (27 a and b):

$$\begin{cases} G^{(1)} = \dfrac{q}{\varepsilon_1\sqrt{x^2+(y-d)^2+z^2}} + \dfrac{\varepsilon_1-\varepsilon_2}{\varepsilon_1+\varepsilon_2}\dfrac{q}{\varepsilon_1\sqrt{x^2+(y+d)^2+z^2}} \\ G^{(2)} = \dfrac{q}{\varepsilon_2\sqrt{x^2+(y-d)^2+z^2}} + \dfrac{\varepsilon_2-\varepsilon_1}{\varepsilon_1+\varepsilon_2}\dfrac{q}{\varepsilon_2\sqrt{x^2+(y-d)^2+z^2}} \end{cases}$$

$$\text{Or } G = \begin{cases} \frac{q}{\varepsilon_1}\left(\frac{1}{\sqrt{x^2+(y-d)^2+z^2}} + \frac{\varepsilon_1-\varepsilon_2}{\varepsilon_1+\varepsilon_2}\frac{1}{\sqrt{x^2+(y+d)^2+z^2}}\right), y > 0 \\ \frac{2\,q}{(\varepsilon_1+\varepsilon_2)\sqrt{x^2+(y-d)^2+z^2}}, y < 0 \end{cases} \tag{30}$$

If we transfer the coordinate system, finally we can obtain the Green`s function:

$$G(x,\tilde{x},y,\tilde{y},z,\tilde{z})=\begin{cases}\frac{q}{\varepsilon_1}\left(\frac{1}{\sqrt{(x-\tilde{x})^2+(y-\tilde{y})^2+(z-\tilde{z})^2}}+\frac{\varepsilon_1-\varepsilon_2}{\varepsilon_1+\varepsilon_2}\frac{1}{\sqrt{(x-\tilde{x})^2+(y+\tilde{y})^2+(z-\tilde{z})^2}}\right), y>0\\ \frac{2\,q}{(\varepsilon_1+\varepsilon_2)\sqrt{(x-\tilde{x})^2+(y-\tilde{y})^2+(z-\tilde{z})^2}}, y<0\end{cases} \quad (31)$$

Let us apply formula (26) now:

$$\varphi=\iiint_{-\infty}^{+\infty}G(x,\tilde{x},y,\tilde{y},z,\tilde{z})\ \frac{1}{v}e^{i\omega\frac{\tilde{x}}{v}}\delta\ (\tilde{y}-d)\delta(\tilde{z})\ d\tilde{x}\ d\tilde{y}\ d\tilde{z}=\int_{-\infty}^{+\infty}G(x,\tilde{x},y,d,z,0)\ \frac{1}{v}e^{i\omega\frac{\tilde{x}}{v}}d\tilde{x}=$$

$$\begin{cases}\int_{-\infty}^{+\infty}\frac{q}{\varepsilon_1}\left(\frac{1}{\sqrt{(x-\tilde{x})^2+(y-d)^2+(z)^2}}+\frac{\varepsilon_1-\varepsilon_2(\omega)}{\varepsilon_1+\varepsilon_2(\omega)}\frac{1}{\sqrt{(x-\tilde{x})^2+(y+d)^2+(z)^2}}\right)\frac{1}{v}e^{i\omega\frac{\tilde{x}}{v}}d\tilde{x},\quad y>0\\ \int_{-\infty}^{+\infty}\frac{2\,q}{(\varepsilon_1+\varepsilon_2(\omega))\sqrt{(x-\tilde{x})^2+(y-d)^2+z^2}}\ \frac{1}{v}e^{i\omega\frac{\tilde{x}}{v}}d\tilde{x},\quad y<0\end{cases} \quad (32)$$

Let us evaluate the potential by transforming (32) to integrals, similar to (21)

$\int_0^{\infty}\frac{cos(k\,ב)}{\sqrt{k^2+\vartheta^2}}\ e^{-א\sqrt{k^2+\vartheta^2}}dk=K_0\left(\vartheta\ \sqrt{ב^2+א^2}\right)$ For such transformation, let us do the following variables interchange: $x-\tilde{x}=\mathfrak{E}$. So, our integrals can be written in the following form:

$$\int_{-\infty}^{+\infty}\frac{q}{A(\omega)\sqrt{\mathfrak{E}^2+\vartheta^2}}e^{i\omega\frac{x}{v}}e^{i\omega\frac{\mathfrak{E}}{v}}d\mathfrak{E}=\int_{-\infty}^{+\infty}\frac{q\left[\cos\left(\omega\frac{\mathfrak{E}}{v}\right)+i\sin\left(\omega\frac{\mathfrak{E}}{v}\right)\right]}{A(\omega)\sqrt{\mathfrak{E}^2+\vartheta^2}}e^{i\omega\frac{x}{v}}d\mathfrak{E}=2\int_0^{+\infty}\frac{q\ \cos\left(\omega\frac{\mathfrak{E}}{v}\right)}{A(\omega)\sqrt{\mathfrak{E}^2+\vartheta^2}}e^{i\omega\frac{x}{v}}d\mathfrak{E}=$$

$$\frac{2\,q}{A(\omega)}e^{i\omega\frac{x}{v}}K_0\left(\vartheta\ \frac{|\omega|}{v}\right) \quad (33)$$

Using (33), one can easily obtain, that

$$\varphi(\vec{r},\omega)=$$

$$\begin{cases}\frac{2\ q}{\varepsilon_1 v}e^{i\omega\frac{x}{v}}\left[K_0\left(\frac{|\omega|}{v}\sqrt{(y-d)^2+(z)^2}\right)+\frac{\varepsilon_1-\varepsilon_2(\omega)}{\varepsilon_1+\varepsilon_2(\omega)}\ K_0\left(\frac{|\omega|}{v}\sqrt{(y+d)^2+(z)^2}\right)\right], y>0\\ \frac{4\,q}{(\varepsilon_1+\varepsilon_2(\omega))v}e^{i\omega\frac{x}{v}}K_0\left(\frac{|\omega|}{v}\sqrt{(y-d)^2+(z)^2}\right), y<0\end{cases} \quad (34)$$

As one can easily see, expression (34) is identical to the expression (25). Now let us do the reverse Fourier transformation from $\omega$ to *t*:

$$\varphi(\vec{r},t)=\frac{1}{2\,\pi}\int_{-\infty}^{+\infty}e^{-i\omega t}\varphi\ (\vec{r},\omega)d\omega \quad (34a)$$

Let us leave aside the functions $K_0$, and consider the expressions with the permittivitys a little closer:

$$\frac{\varepsilon_1-\varepsilon_2(\omega)}{\varepsilon_1+\varepsilon_2(\omega)}=\frac{\varepsilon_1\left(\omega_{to}^2-\omega^2-2i\gamma\omega\right)-\chi_\infty\left(\omega_{lo}^2-\omega^2-2i\gamma\omega\right)}{\varepsilon_1\left(\omega_{to}^2-\omega^2-2i\gamma\omega\right)+\chi_\infty\left(\omega_{lo}^2-\omega^2-2i\gamma\omega\right)}=1-2\frac{\chi_\infty\left(\omega_{lo}^2-\omega^2-2i\gamma\omega\right)}{\varepsilon_1\left(\omega_{to}^2-\omega^2-2i\gamma\omega\right)+\chi_\infty\left(\omega_{lo}^2-\omega^2-2i\gamma\omega\right)} \quad (35a)$$

$$\frac{1}{\varepsilon_1+\varepsilon_2(\omega)}=\frac{\omega_{to}^2-\omega^2-2i\gamma\omega}{\varepsilon_1\left(\omega_{to}^2-\omega^2-2i\gamma\omega\right)+\chi_\infty\left(\omega_{lo}^2-\omega^2-2i\gamma\omega\right)}=\frac{1}{\varepsilon_1}\left[1-\frac{\chi_\infty\left(\omega_{lo}^2-\omega^2-2i\gamma\omega\right)}{\varepsilon_1\left(\omega_{to}^2-\omega^2-2i\gamma\omega\right)+\chi_\infty\left(\omega_{lo}^2-\omega^2-2i\gamma\omega\right)}\right]=\frac{1}{\varepsilon_1}-\frac{\frac{\chi_\infty}{\varepsilon_1}\left(\omega_{lo}^2-\omega^2-2i\gamma\omega\right)}{\varepsilon_1\left(\omega_{to}^2-\omega^2-2i\gamma\omega\right)+\chi_\infty\left(\omega_{lo}^2-\omega^2-2i\gamma\omega\right)} \quad (35b)$$

We have separated frequency-dependent and nondependent terms in these expressions. The nondependent ones lead to the transformation of (34) into the well-known potentials of mirror image with coordinate shift $x \to x - vt$:

$$\frac{1}{2\pi}\int_{-\infty}^{+\infty}\frac{2\,q}{\varepsilon_1 v}e^{-i\omega\left(t-\frac{x}{v}\right)}\left[K_0\left(\frac{|\omega|}{v}\sqrt{(y-d)^2+(z)^2}\right)+\ K_0\left(\frac{|\omega|}{v}\sqrt{(y+d)^2+(z)^2}\right)\right]d\omega =$$
$$\frac{1}{\pi}\int_{-\infty}^{+\infty}\frac{q}{\varepsilon_1 v}cos\left(\omega\left(t-\frac{x}{v}\right)\right)\left[K_0\left(\frac{|\omega|}{v}\sqrt{(y-d)^2+(z)^2}\right)+\ K_0\left(\frac{|\omega|}{v}\sqrt{(y+d)^2+(z)^2}\right)\right]d\omega \tag{36a}$$

$$\frac{1}{2\pi}\int_{-\infty}^{+\infty}\frac{4\,q}{\varepsilon_1 v}e^{-i\omega\left(t-\frac{x}{v}\right)}\ K_0\left(\frac{|\omega|}{v}\sqrt{(y-d)^2+(z)^2}\right)d\omega =$$
$$\frac{1}{\pi}\int_{-\infty}^{+\infty}\frac{2\,q}{\varepsilon_1 v}cos\left(\omega\left(t-\frac{x}{v}\right)\right)\ K_0\left(\frac{|\omega|}{v}\sqrt{(y-d)^2+(z)^2}\right)d\omega \tag{36b}$$

Now let us consider the frequency-dependent terms in (34a):

$$\frac{1}{2\pi}\int_{-\infty}^{+\infty}\frac{2\,q}{\varepsilon_1 v}e^{-i\omega\left(t-\frac{x}{v}\right)}\left[K_0\left(\frac{|\omega|}{v}\sqrt{(y-d)^2+(z)^2}\right)-\right.$$
$$\left.2\frac{\chi_\infty(\omega_{lo}^2-\omega^2-2i\gamma\omega)}{\varepsilon_1(\omega_{to}^2-\omega^2-2i\gamma\omega)+\chi_\infty(\omega_{lo}^2-\omega^2-2i\gamma\omega)}\ K_0\left(\frac{|\omega|}{v}\sqrt{(y+d)^2+(z)^2}\right)\right]d\omega = \left|\begin{matrix}\omega'=\frac{\omega}{v}\\ d\omega'=\frac{d\omega}{v}\end{matrix}\right| =$$
$$\frac{q}{\pi\,\varepsilon_1}\int_{-\infty}^{+\infty}e^{-i\omega'(vt-x)}\left[K_0\left(|\omega'|\sqrt{(y-d)^2+(z)^2}\right)-\right.$$
$$\left.2\frac{\chi_\infty(\omega_{lo}^2-(\omega' v)^2-2i\gamma\omega' v)}{\varepsilon_1(\omega_{to}^2-(\omega' v)^2-2i\gamma\omega' v)+\chi_\infty(\omega_{lo}^2-(\omega' v)^2-2i\gamma\omega' v)}\ K_0\left(|\omega'|\sqrt{(y+d)^2+(z)^2}\right)\right]d\omega' \tag{37}$$

For less cumbersome expressions, let us introduce such values:

$$\begin{gathered}A=\frac{q}{\pi\,\varepsilon_1}\\ \tilde{x}=vt-x\\ B_\pm=\sqrt{(y\pm d)^2+(z)^2}\end{gathered} \tag{37a}$$

Using (37a), we are facing the following integral:

$$I = A\int_{-\infty}^{+\infty}e^{-i\omega'\tilde{x}}\left[K_0(|\omega'|\,B_-)-2\frac{\chi_\infty(\omega_{lo}^2-(\omega' v)^2-2i\gamma\omega' v)}{\varepsilon_1(\omega_{to}^2-(\omega' v)^2-2i\gamma\omega' v)+\chi_\infty(\omega_{lo}^2-(\omega' v)^2-2i\gamma\omega' v)}\ K_0(|\omega'|\,B_+)\right]d\omega' \tag{37b}$$

$$\frac{\chi_\infty(\omega_{lo}^2-(\omega' v)^2-2i\gamma\omega' v)}{\varepsilon_1(\omega_{to}^2-(\omega' v)^2-2i\gamma\omega' v)+\chi_\infty(\omega_{lo}^2-(\omega' v)^2-2i\gamma\omega' v)} =$$
$$\frac{\chi_\infty{}^2\left(\left(\omega_{lo}^2-(\omega' v)^2\right)^2+4\left(\gamma\omega' v\right)^2\right)+\chi_\infty\varepsilon_1(\omega_{lo}^2-(\omega' v)^2-2i\gamma\omega' v)\left(\omega_{to}^2-(\omega' v)^2+2i\gamma\omega' v\right)}{\left(\varepsilon_1(\omega_{to}^2-(\omega' v)^2)+\chi_\infty(\omega_{lo}^2-(\omega' v)^2)\right)^2+4(\gamma\omega' v)^2(\varepsilon_1+\chi_\infty)^2} =$$
$$\frac{\chi_\infty{}^2\left(\left(\omega_{lo}^2-(\omega' v)^2\right)^2+4\left(\gamma\omega' v\right)^2\right)+\chi_\infty\varepsilon_1\left((\omega_{lo}^2-(\omega' v)^2)(\omega_{to}^2-(\omega' v)^2)+4\left(\gamma\omega' v\right)^2\right)}{\left(\varepsilon_1(\omega_{to}^2-(\omega' v)^2)+\chi_\infty(\omega_{lo}^2-(\omega' v)^2)\right)^2+4(\gamma\omega' v)^2(\varepsilon_1+\chi_\infty)^2} +$$
$$i\frac{-2\,\chi_\infty\varepsilon_1\gamma\omega' v(\omega_{to}^2-\omega_{lo}^2)}{\left(\varepsilon_1(\omega_{to}^2-(\omega' v)^2)+\chi_\infty(\omega_{lo}^2-(\omega' v)^2)\right)^2+4(\gamma\omega' v)^2(\varepsilon_1+\chi_\infty)^2} \tag{37c}$$

So, using (37c), we can separate the real and imaginary parts in (37b). Also it is important to take into account that the integral of an odd function by symmetric limits equals to zero:

$$Re(I) = I_1 = A\int_{-\infty}^{\infty}\Bigg[\cos(\omega'\tilde{x})\,K_0(|\omega'|\,B_-) - 2\left(\cos(\omega'\tilde{x})\frac{\chi_\infty{}^2\left(\left(\omega_{lo}^2-(\omega' v)^2\right)^2+4\left(\gamma\omega' v\right)^2\right)+\chi_\infty\varepsilon_1\left(\left(\omega_{lo}^2-(\omega' v)^2\right)\left(\omega_{to}^2-(\omega' v)^2\right)+4\left(\gamma\omega' v\right)^2\right)}{\left(\varepsilon_1\left(\omega_{to}^2-(\omega' v)^2\right)+\chi_\infty\left(\omega_{lo}^2-(\omega' v)^2\right)\right)^2+4(\gamma\omega' v)^2(\varepsilon_1+\chi_\infty)^2} + sin(\omega'\tilde{x})\frac{2\,\chi_\infty\varepsilon_1\gamma\omega' v\left(\omega_{to}^2-\omega_{lo}^2\right)}{\left(\varepsilon_1\left(\omega_{to}^2-(\omega' v)^2\right)+\chi_\infty\left(\omega_{lo}^2-(\omega' v)^2\right)\right)^2+4(\gamma\omega' v)^2(\varepsilon_1+\chi_\infty)^2}\right)K_0(|\omega'|\,B_+)\Bigg]d\omega' \quad (37d)$$

$$Im(I) = A\int_{-\infty}^{\infty}\Bigg[-\sin(\omega'\tilde{x})\,K_0(|\omega'|\,B_-) + 2\left(\sin(\omega'\tilde{x})\frac{\chi_\infty{}^2\left(\left(\omega_{lo}^2-(\omega' v)^2\right)^2+4\left(\gamma\omega' v\right)^2\right)+\chi_\infty\varepsilon_1\left(\left(\omega_{lo}^2-(\omega' v)^2\right)\left(\omega_{to}^2-(\omega' v)^2\right)+4\left(\gamma\omega' v\right)^2\right)}{\left(\varepsilon_1\left(\omega_{to}^2-(\omega' v)^2\right)+\chi_\infty\left(\omega_{lo}^2-(\omega' v)^2\right)\right)^2+4(\gamma\omega' v)^2(\varepsilon_1+\chi_\infty)^2} + cos(\omega'\tilde{x})\frac{2\,\chi_\infty\varepsilon_1\gamma\omega' v\left(\omega_{to}^2-\omega_{lo}^2\right)}{\left(\varepsilon_1\left(\omega_{to}^2-(\omega' v)^2\right)+\chi_\infty\left(\omega_{lo}^2-(\omega' v)^2\right)\right)^2+4(\gamma\omega' v)^2(\varepsilon_1+\chi_\infty)^2}\right)K_0(|\omega'|\,B_+)\Bigg]d\omega' = 0 \quad (37e)$$

Let us nondimensionalaze (37d):

$$2A\int_0^{\infty} d\omega'[-2\cos(\omega'\tilde{x})\,\tau(\omega')\,K_0(\omega'\,B_+)] \quad (38a)$$

$$\tau(\omega') = \frac{\chi_\infty{}^2\left(\left(\omega_{lo}^2-(\omega' v)^2\right)^2+4\left(\gamma\omega' v\right)^2\right)+\chi_\infty\varepsilon_1\left(\left(\omega_{lo}^2-(\omega' v)^2\right)\left(\omega_{to}^2-(\omega' v)^2\right)+4\left(\gamma\omega' v\right)^2\right)}{\left(\varepsilon_1\left(\omega_{to}^2-(\omega' v)^2\right)+\chi_\infty\left(\omega_{lo}^2-(\omega' v)^2\right)\right)^2+4(\gamma\omega' v)^2(\varepsilon_1+\chi_\infty)^2} \quad (38b)$$

In (38b), we will assume that $\varepsilon_1 = 1$:

$$\frac{\chi_\infty{}^2\left(\left(\omega_{lo}^2-\omega^2\right)^2+4(\gamma\omega)^2\right)+\chi_\infty\left(\left(\omega_{lo}^2-(\omega)^2\right)\left(\omega_{to}^2-(\omega)^2\right)+4(\gamma\omega)^2\right)}{\left(\left(\omega_{to}^2-\omega^2\right)+\chi_\infty\left(\omega_{lo}^2-\omega^2\right)\right)^2+4(\gamma\omega)^2(1+\chi_\infty)^2} \quad (39)$$

Let us transform $B_\pm$:

$$\begin{gathered} B_\pm=\sqrt{(y\pm d)^2+(z)^2}=d\sqrt{(\acute{y}\pm 1)^2+(\acute{z})^2} \\ B_\pm=d\tilde{B}_\pm \\ \tilde{B}_\pm=\sqrt{(\acute{y}\pm 1)^2+(\acute{z})^2} \\ \tilde{x}=x-vt \\ \acute{y}=\frac{y}{d} \\ \acute{z}=\frac{z}{d} \end{gathered} \quad (40)$$

So, after we applied (40), the MacDonald's function in (38a) has such argument:

$$\omega'\,B_\pm = \frac{\omega}{v}d\tilde{B}_\pm = \left|\begin{matrix}\frac{\omega}{v}d=\frac{\omega}{\omega_v}=\widetilde{\omega} \\ \frac{v}{d}=\omega_v\end{matrix}\right| = \widetilde{\omega}\tilde{B}_\pm \quad (41)$$

$$\omega'\tilde{x} = \frac{\omega}{v}d\,\frac{x-vt}{d} = \widetilde{\omega}\acute{x} \quad (42)$$

So, (38a) can be re-written in a following way:

$$2\frac{A}{d}\int_0^{\infty} d\widetilde{\omega}\left[\cos(\widetilde{\omega}\acute{x})\,K_0\left(\widetilde{\omega}\tilde{B}_-\right) - 2\cos(\widetilde{\omega}\acute{x})\,\tau(\widetilde{\omega}\,\omega_v)\,K_0\left(\widetilde{\omega}\tilde{B}_+\right)\right]d\widetilde{\omega} \quad (43)$$

Now it is possible to nondimensionalize the expression (39):

$$\frac{\chi_\infty{}^2\left(\left(\omega_{lo}^2-\omega^2\right)^2+4(\gamma\omega)^2\right)+\chi_\infty\left(\left(\omega_{lo}^2-(\omega)^2\right)\left(\omega_{to}^2-(\omega)^2\right)+4(\gamma\omega)^2\right)}{\left(\left(\omega_{to}^2-\omega^2\right)+\chi_\infty\left(\omega_{lo}^2-\omega^2\right)\right)^2+4(\gamma\omega)^2(1+\chi_\infty)^2}=$$

$$\frac{\chi_\infty{}^2\left(\left(\tilde{\omega}_{lo}^2-\tilde{\omega}^2\right)^2+4(\tilde{\gamma}\tilde{\omega})^2\right)+\chi_\infty\left(\left(\tilde{\omega}_{lo}^2-(\tilde{\omega})^2\right)\left(\tilde{\omega}_{to}^2-(\tilde{\omega})^2\right)+4(\tilde{\gamma}\tilde{\omega})^2\right)}{\left(\left(\tilde{\omega}_{to}^2-\tilde{\omega}^2\right)+\chi_\infty\left(\tilde{\omega}_{lo}^2-\tilde{\omega}^2\right)\right)^2+4(\tilde{\gamma}\tilde{\omega})^2(1+\chi_\infty)^2}=\tau \tag{44}$$

In (44) the following nondimensional parameters were introduced:

$$\tilde{\gamma}=\frac{\gamma}{\omega_v}$$
$$\tilde{\omega}_{lo}=\frac{\omega_{lo}}{\omega_v} \tag{45}$$
$$\tilde{\omega}_{to}=\frac{\omega_{to}}{\omega_v}$$

Finally, using (40), (41) and (45) our integral can be re-written in the following form:

$$2\frac{q}{\pi d}\int_0^\infty d\tilde{\omega}\left[-2\left(\cos(\tilde{\omega}\acute{x})\frac{\chi_\infty{}^2\left(\left(\tilde{\omega}_{lo}^2-\tilde{\omega}^2\right)^2+4(\tilde{\gamma}\tilde{\omega})^2\right)+\chi_\infty\left(\left(\tilde{\omega}_{lo}^2-(\tilde{\omega})^2\right)\left(\tilde{\omega}_{to}^2-(\tilde{\omega})^2\right)+4(\tilde{\gamma}\tilde{\omega})^2\right)}{\left(\left(\tilde{\omega}_{to}^2-\tilde{\omega}^2\right)+\chi_\infty\left(\tilde{\omega}_{lo}^2-\tilde{\omega}^2\right)\right)^2+4(\tilde{\gamma}\tilde{\omega})^2(1+\chi_\infty)^2}+\right.\right.$$

$$\left.\left.sin(\tilde{\omega}\acute{x})\frac{2\chi_\infty\tilde{\gamma}\tilde{\omega}\left(\tilde{\omega}_{to}^2-\tilde{\omega}_{lo}^2\right)}{\left(\left(\tilde{\omega}_{to}^2-\tilde{\omega}^2\right)+\chi_\infty\left(\tilde{\omega}_{lo}^2-\tilde{\omega}^2\right)\right)^2+4(\tilde{\gamma}\tilde{\omega})^2(1+\chi_\infty)^2}\right)K_0\left(\tilde{\omega}\tilde{B}_+\right)\right]d\tilde{\omega} \tag{46}$$

Now, let us consider the potential in the lower half-space:

$$\varphi(\vec{r},\omega)=\frac{4\,q}{(\varepsilon_1+\varepsilon_2(\omega))v}e^{i\omega\frac{x}{v}}K_0\left(\frac{|\omega|}{v}\sqrt{(y-d)^2+(z)^2}\right),y<0 \tag{47}$$

And

$$\frac{1}{\varepsilon_1+\varepsilon_2(\omega)}=\frac{1}{\varepsilon_1}-\frac{\frac{\chi_\infty}{\varepsilon_1}\left(\omega_{lo}^2-\omega^2-2i\gamma\omega\right)}{\varepsilon_1\left(\omega_{to}^2-\omega^2-2i\gamma\omega\right)+\chi_\infty\left(\omega_{lo}^2-\omega^2-2i\gamma\omega\right)} \tag{48}$$

Also when we were transforming (32) we introduced such nondimensional variables and parameters:

$$x'=\frac{x-vt}{d}$$
$$y'=\frac{y}{d} \tag{49}$$
$$z'=\frac{z}{d}$$

$$\tilde{\omega}=\frac{\omega}{v}d$$
$$\omega_v=\frac{v}{d} \tag{50}$$

$$\tilde{\gamma}=\frac{\gamma}{\omega_v}$$
$$\tilde{\omega}_{lo}=\frac{\omega_{lo}}{\omega_v} \tag{51}$$
$$\tilde{\omega}_{to}=\frac{\omega_{to}}{\omega_v}$$

Now let us imply (49-51) in the reverse Fourier transformation of (47):

$$I=-\frac{4\;q}{2\,\pi\,v}\frac{\chi_\infty}{\varepsilon_1}\int_{-\infty}^{+\infty}e^{-i\omega\left(t-\frac{x}{v}\right)}\frac{\left(\omega_{lo}^2-\omega^2-2i\gamma\omega\right)}{\varepsilon_1\left(\omega_{to}^2-\omega^2-2i\gamma\omega\right)+\chi_\infty\left(\omega_{lo}^2-\omega^2-2i\gamma\omega\right)}\;K_0\left(\frac{|\omega|}{v}\sqrt{(y-d)^2+(z)^2}\right)d\omega \tag{52}$$

The most mind-defying coefficient in (52) can be transformed in a following way:

$$Im(\gamma) = Im\left(\frac{(\omega_{lo}^2-\omega^2-2i\gamma\omega)}{\varepsilon_1(\omega_{to}^2-\omega^2-2i\gamma\omega)+\chi_\infty(\omega_{lo}^2-\omega^2-2i\gamma\omega)}\right) = \frac{-2\gamma\omega\varepsilon_1(\omega_{to}^2-\omega_{lo}{}^2)}{[\varepsilon_1(\omega_{to}^2-\omega^2)+\chi_\infty(\omega_{lo}^2-\omega^2)]^2+4\gamma^2\omega^2(\varepsilon_1+\chi_\infty)^2} \quad (53)$$

$$Re(\gamma) = Re\left(\frac{(\omega_{lo}^2-\omega^2-2i\gamma\omega)}{\varepsilon_1(\omega_{to}^2-\omega^2-2i\gamma\omega)+\chi_\infty(\omega_{lo}^2-\omega^2-2i\gamma\omega)}\right) = \frac{\chi_\infty\left(\left(\omega_{lo}^2-\omega^2\right)^2+4(\gamma\omega)^2\right)+\varepsilon_1\left(\left(\omega_{lo}^2-\omega^2\right)(\omega_{to}^2-\omega^2)+4(\gamma\omega)^2\right)}{[\varepsilon_1(\omega_{to}^2-\omega^2)+\chi_\infty(\omega_{lo}^2-\omega^2)]^2+4\gamma^2\omega^2(\varepsilon_1+\chi_\infty)^2} \quad (54)$$

So, when we consider that (52) includes a complex exponent, we obtain the following expressions:

$$Im(I) \equiv 0 \quad (55)$$

$$Re(I) = -\frac{4}{\pi}\frac{q}{v}\frac{\chi_\infty}{\varepsilon_1}\int_0^{+\infty} d\omega\left\{cos\left(\omega\frac{vt-x}{v}\right)\frac{\chi_\infty\left(\left(\omega_{lo}^2-\omega^2\right)^2+4(\gamma\omega)^2\right)+\varepsilon_1\left(\left(\omega_{lo}^2-\omega^2\right)(\omega_{to}^2-\omega^2)+4(\gamma\omega)^2\right)}{[\varepsilon_1(\omega_{to}^2-\omega^2)+\chi_\infty(\omega_{lo}^2-\omega^2)]^2+4\gamma^2\omega^2(\varepsilon_1+\chi_\infty)^2} + sin\left(\omega\frac{vt-x}{v}\right)\frac{2\gamma\omega\varepsilon_1(\omega_{to}^2-\omega_{lo}{}^2)}{[\varepsilon_1(\omega_{to}^2-\omega^2)+\chi_\infty(\omega_{lo}^2-\omega^2)]^2+4\gamma^2\omega^2(\varepsilon_1+\chi_\infty)^2}\right\} K_0\left(\frac{|\omega|}{v}\sqrt{(y-d)^2+(z)^2}\right) \quad (56)$$

Note that the integrated expression in (56) is an even function of $\omega$.

$$\text{Let } \tilde{A} = -\frac{4}{\pi}\frac{v\,q}{}\frac{\chi_\infty}{\varepsilon_1} \quad (56a)$$

$$\text{Also } \frac{\omega}{v}\sqrt{(y-d)^2+(z)^2} = \tilde{\omega}\tilde{B}_- \quad (56b)$$

$$\omega\frac{vt-x}{v} = x'\tilde{\omega} \quad (56c)$$

Then one can obtain that (56) can be re-written in the following form:

$$Re(I) = -\frac{4}{\pi}\frac{q}{d}\frac{\chi_\infty}{\varepsilon_1}\int_0^{+\infty} d\,\tilde{\omega}\left\{cos(x'\tilde{\omega})\frac{\chi_\infty\left(\left(\tilde{\omega}_{lo}^2-\tilde{\omega}^2\right)^2+4(\tilde{\gamma}\tilde{\omega})^2\right)+\varepsilon_1\left(\left(\tilde{\omega}_{lo}^2-\tilde{\omega}^2\right)(\tilde{\omega}_{to}^2-\tilde{\omega}^2)+4(\tilde{\gamma}\tilde{\omega})^2\right)}{[\varepsilon_1(\tilde{\omega}_{to}^2-\tilde{\omega}^2)+\chi_\infty(\tilde{\omega}_{lo}^2-\tilde{\omega}^2)]^2+4\tilde{\gamma}^2\tilde{\omega}^2(\varepsilon_1+\chi_\infty)^2} + sin(x'\tilde{\omega})\frac{2\tilde{\gamma}\varepsilon_1\tilde{\omega}(\tilde{\omega}_{to}^2-\tilde{\omega}_{lo}{}^2)}{[\varepsilon_1(\tilde{\omega}_{to}^2-\tilde{\omega}^2)+\chi_\infty(\tilde{\omega}_{lo}^2-\tilde{\omega}^2)]^2+4\tilde{\gamma}^2\tilde{\omega}^2(\varepsilon_1+\chi_\infty)^2}\right\} K_0(\tilde{\omega}\tilde{B}_-) \quad (57)$$

In (57) $\tilde{\varepsilon}_1 = \frac{\varepsilon_1}{\omega_v}$. Now we can finally leave the burden of calculating the integral (57) to Wolfram Mathematica.

Let us find the energy losses in a dielectric media:

$$Q = -\frac{1}{4\pi}\int_{-\infty}^{+\infty} dt \iiint_V \vec{E}\frac{\partial\vec{D}}{\partial t}dV \quad (58)$$

In (58) we can do a Fourier transform:

$$\vec{D}(\vec{r},\omega) = \varepsilon(\omega)\,\vec{E}(\vec{r},\omega) \quad (59)$$

So, the integrand in (58) can be written in the following form:

$$\rho_Q = -\frac{1}{4\pi}\int_{-\infty}^{+\infty}\vec{E}(\vec{r},\omega')e^{-i\omega' t}d\omega'\int_{-\infty}^{+\infty}\varepsilon(\omega)\frac{\partial\vec{E}(\vec{r},\omega)}{\partial t}\,e^{-i\omega t}d\omega \quad (60)$$

In (60) $\vec{E}(\vec{r},\omega') = -\vec{\nabla}\varphi(\vec{r},\omega')$ (60a)

Also

$$Re\left[\varphi(\vec{r},\omega)\, e^{-i\omega t}\right] = -\frac{2}{\pi}\frac{q}{d}\frac{\chi_\infty}{\varepsilon_1}\left\{cos\left(x'\,\tilde{\omega}\right)\tau(\omega) + sin\left(x'\,\tilde{\omega}\right)\sigma(\omega)\right\} K_0\left(|\tilde{\omega}|\tilde{B}_-\right) \quad (61a)$$

$$Im\left[\varphi(\vec{r},\omega)\, e^{-i\omega t}\right] = -\frac{2}{\pi}\frac{q}{d}\frac{\chi_\infty}{\varepsilon_1}\left\{-cos\left(x'\,\tilde{\omega}\right)\sigma(\omega) + sin\left(x'\,\tilde{\omega}\right)\tau(\omega)\right\} K_0\left(|\tilde{\omega}|\tilde{B}_-\right) \quad (61b)$$

$$\frac{\omega}{v}\sqrt{(y-d)^2+z^2} = \tilde{\omega}\tilde{B}_- \quad (61c)$$

$$x' = \frac{x-v\,t}{d} \quad (61d)$$

$$\begin{cases} \frac{\partial}{\partial y} K_0\left(\tilde{\omega}\tilde{B}_-\right) = -\frac{\tilde{\omega}\left(y'-1\right)}{\tilde{B}_-} K_1\left(\tilde{\omega}\tilde{B}_-\right) \\ \frac{\partial}{\partial z} K_0\left(\tilde{\omega}\tilde{B}_-\right) = -\frac{\tilde{\omega}\, z'}{\tilde{B}_-} K_1\left(\tilde{\omega}\tilde{B}_-\right) \end{cases} \quad (61e)$$

$$\tau(\omega) = \frac{\chi_\infty\left(\left(\tilde{\omega}_{lo}^2-\tilde{\omega}^2\right)^2+4(\tilde{\gamma}\tilde{\omega})^2\right)+\varepsilon_1\left(\left(\tilde{\omega}_{lo}^2-\tilde{\omega}^2\right)\left(\tilde{\omega}_{to}^2-\tilde{\omega}^2\right)+4(\tilde{\gamma}\tilde{\omega})^2\right)}{\left[\varepsilon_1\left(\tilde{\omega}_{to}^2-\tilde{\omega}^2\right)+\chi_\infty\left(\tilde{\omega}_{lo}^2-\tilde{\omega}^2\right)\right]^2+4\,\tilde{\gamma}^2\tilde{\omega}^2\left(\varepsilon_1+\chi_\infty\right)^2} \quad (61f)$$

$$\sigma(\omega) = \frac{2\tilde{\gamma}\varepsilon_1\tilde{\omega}\left(\tilde{\omega}_{to}^2-\tilde{\omega}_{lo}{}^2\right)}{\left[\varepsilon_1\left(\tilde{\omega}_{to}^2-\tilde{\omega}^2\right)+\chi_\infty\left(\tilde{\omega}_{lo}^2-\tilde{\omega}^2\right)\right]^2+4\,\tilde{\gamma}^2\tilde{\omega}^2\left(\varepsilon_1+\chi_\infty\right)^2} \quad (61g)$$

And $\varepsilon(\omega) = \chi_\infty\frac{\omega_{lo}^2-\omega^2-2i\gamma\omega}{\omega_{to}^2-\omega^2-2i\gamma\omega}$, so

$$Im(\varepsilon(\omega)\,) = \chi_\infty\frac{2\gamma\omega\left(\omega_{lo}^2-\omega_{to}^2\right)}{\left(\omega_{to}^2-\omega^2\right)^2+4\gamma^2\omega^2} \quad (62a)$$

$$Re(\varepsilon(\omega)\,) = \chi_\infty\frac{\left(\omega_{to}^2-\omega^2\right)\left(\omega_{lo}^2-\omega^2\right)+4\gamma^2\omega^2}{\left(\omega_{to}^2-\omega^2\right)^2+4\gamma^2\omega^2} \quad (62b)$$

Let us find the electric field and its induction in the dispersive media:

$$\vec{E} = \frac{4}{\pi}\frac{q}{d}\frac{\chi_\infty}{\varepsilon_1}\vec{\nabla}\int_0^{+\infty} d\,\tilde{\omega}\{cos(x'\tilde{\omega})\tau(\omega) + sin(x'\tilde{\omega})\sigma(\omega)\}\, K_0\left(\tilde{\omega}\tilde{B}_-\right) \quad (63)$$

Or

$$E_x = \frac{4}{\pi}\frac{q}{d^2}\frac{\chi_\infty}{\varepsilon_1}\int_0^{+\infty}\tilde{\omega}\{-sin(x'\tilde{\omega})\tau(\omega) + cos(x'\tilde{\omega})\sigma(\omega)\}\, K_0\left(\tilde{\omega}\tilde{B}_-\right) d\tilde{\omega} \quad (64a)$$

$$E_y = -\frac{4}{\pi}\frac{q}{d^2}\frac{\chi_\infty}{\varepsilon_1}\int_0^{+\infty}\{cos(x'\tilde{\omega})\tau(\omega) + sin(x'\tilde{\omega})\sigma(\omega)\}\frac{\tilde{\omega}\,(y'-1)}{\tilde{B}_-}K_1\left(\tilde{\omega}\tilde{B}_-\right) d\tilde{\omega} \quad (64b)$$

$$E_z = -\frac{4}{\pi}\frac{q}{d^2}\frac{\chi_\infty}{\varepsilon_1}\int_0^{+\infty}\{cos(x'\tilde{\omega})\tau(\omega) + sin(x'\tilde{\omega})\sigma(\omega)\}\frac{\tilde{\omega}\,z'}{\tilde{B}_-}K_1\left(\tilde{\omega}\tilde{B}_-\right) d\tilde{\omega} \quad (64c)$$

Let us find the induction now:

$$\vec{D} = -\vec{\nabla}\int_{-\infty}^{+\infty}\varepsilon(\omega)\varphi_\omega e^{-i\omega t}d\omega \quad (65)$$

We know that the real part of the permittivity is even, the imaginary one is uneven. The same is true for the expression $\varphi_\omega e^{-i\omega t}$. So, we can write as follows:

$$\vec{D} = -\vec{\nabla}\int_0^{+\infty}\left\{\varepsilon'(\omega)Re\left[\varphi(\vec{r},\omega)\,e^{-i\omega t}\right] - \varepsilon''(\omega)Im\left[\varphi(\vec{r},\omega)\,e^{-i\omega t}\right]\right\}d\omega \tag{66}$$

Let us imply (61a) and (61b):

$$\vec{D} = \vec{\nabla}\frac{4}{\pi}\frac{q}{d}\frac{\chi_\infty}{\varepsilon_1}\int_0^{+\infty}\left\{\varepsilon'(\tilde{\omega})\{cos(x'\,\tilde{\omega})\tau(\omega) + sin(x'\,\tilde{\omega})\sigma(\omega)\} - \varepsilon''(\tilde{\omega})\{-cos(x'\,\tilde{\omega})\sigma(\omega) + sin(x'\,\tilde{\omega})\tau(\omega)\}\right\}K_0\left(\tilde{\omega}\tilde{B}_-\right)d\tilde{\omega} \tag{67}$$

Now we can easily apply the gradient operator in (67):

$$D_x = \frac{4}{\pi}\frac{q}{d^2}\frac{\chi_\infty}{\varepsilon_1}\int_0^{+\infty}\tilde{\omega}\left\{\varepsilon'(\tilde{\omega})\{-sin(x'\,\tilde{\omega})\tau(\omega) + cos(x'\,\tilde{\omega})\sigma(\omega)\} - \varepsilon''(\tilde{\omega})\{sin(x'\,\tilde{\omega})\sigma(\omega) + cos(x'\,\tilde{\omega})\tau(\omega)\}\right\}K_0\left(\tilde{\omega}\tilde{B}_-\right)d\tilde{\omega} \tag{68a}$$

$$D_y = -\frac{4}{\pi}\frac{q}{d^2}\frac{\chi_\infty}{\varepsilon_1}\int_0^{+\infty}\left\{\varepsilon'(\tilde{\omega})\{cos(x'\,\tilde{\omega})\tau(\omega) + sin(x'\,\tilde{\omega})\sigma(\omega)\} - \varepsilon''(\tilde{\omega})\{-cos(x'\,\tilde{\omega})\sigma(\omega) + sin(x'\,\tilde{\omega})\tau(\omega)\}\right\}\frac{\tilde{\omega}\,(y'-1)}{\tilde{B}_-}K_1\left(\tilde{\omega}\tilde{B}_-\right)d\tilde{\omega} \tag{68b}$$

$$D_z = -\frac{4}{\pi}\frac{q}{d^2}\frac{\chi_\infty}{\varepsilon_1}\int_0^{+\infty}\left\{\varepsilon'(\tilde{\omega})\{cos(x'\,\tilde{\omega})\tau(\omega) + sin(x'\,\tilde{\omega})\sigma(\omega)\} - \varepsilon''(\tilde{\omega})\{-cos(x'\,\tilde{\omega})\sigma(\omega) + sin(x'\,\tilde{\omega})\tau(\omega)\}\right\}\frac{\tilde{\omega}\,z'}{\tilde{B}_-}K_1\left(\tilde{\omega}\tilde{B}_-\right)d\tilde{\omega} \tag{68c}$$

The connection between the induction and the field is known from classical electrodynamics:

$$\vec{D} = \vec{E} + 4\pi\,\vec{P} \tag{69}$$

So,

$$\vec{P} = \frac{\vec{D}-\vec{E}}{4\,\pi} \tag{70}$$

Or, implying the view of the field and induction from Eqs. (64a-c) and Eqs. (68a-c) we can obtain, that

$$P_x = \frac{q}{\pi^2\,d^2}\frac{\chi_\infty}{\varepsilon_1}\int_0^{+\infty}\tilde{\omega}\left\{[\varepsilon'(\tilde{\omega}) - 1]\{-sin(x'\,\tilde{\omega})\tau(\omega) + cos(x'\,\tilde{\omega})\sigma(\omega)\} - \varepsilon''(\tilde{\omega})\{sin(x'\,\tilde{\omega})\sigma(\omega) + cos(x'\,\tilde{\omega})\tau(\omega)\}\right\}K_0\left(\tilde{\omega}\tilde{B}_-\right)d\tilde{\omega} \tag{71a}$$

$$P_y = -\frac{q}{\pi^2\,d^2}\frac{\chi_\infty}{\varepsilon_1}\int_0^{+\infty}\left\{[\varepsilon'(\tilde{\omega}) - 1]\{cos(x'\,\tilde{\omega})\tau(\omega) + sin(x'\,\tilde{\omega})\sigma(\omega)\} - \varepsilon''(\tilde{\omega})\{-cos(x'\,\tilde{\omega})\sigma(\omega) + sin(x'\,\tilde{\omega})\tau(\omega)\}\right\}\frac{\tilde{\omega}\,(y'-1)}{\tilde{B}_-}K_1\left(\tilde{\omega}\tilde{B}_-\right)d\tilde{\omega} \tag{71b}$$

$$P_z = -\frac{q}{\pi^2\,d^2}\frac{\chi_\infty}{\varepsilon_1}\int_0^{+\infty}\left\{[\varepsilon'(\tilde{\omega}) - 1]\{cos(x'\,\tilde{\omega})\tau(\omega) + sin(x'\,\tilde{\omega})\sigma(\omega)\} - \varepsilon''(\tilde{\omega})\{-cos(x'\,\tilde{\omega})\sigma(\omega) + sin(x'\,\tilde{\omega})\tau(\omega)\}\right\}\frac{\tilde{\omega}\,z'}{\tilde{B}_-}K_1\left(\tilde{\omega}\tilde{B}_-\right)d\tilde{\omega} \tag{71c}$$

All of these formulas were derived omitting the non-dispersive terms.